\documentclass[aps,prb,twocolumn,preprintnumbers,amsmath,amssymb,superscriptaddress,floatfix]{revtex4}
\usepackage{graphicx, bm}
\usepackage{latexsym}
\usepackage{amsmath}
\usepackage{graphics}
\usepackage{amssymb}
\usepackage{layout}
\usepackage{verbatim}
\usepackage{amsfonts,epsfig}
\newcommand{\ket}[1]{\left|#1\right>}

\newcommand{\beq}{\begin{equation}}
\newcommand{\eeq}{\end{equation}}

\begin{document}
\title{Exchange coupling in silicon quantum dots: Theoretical considerations for quantum computation}
\author{Qiuzi Li}
\affiliation{Condensed Matter Theory Center, Department of Physics, University of Maryland, College Park, Maryland 20742-4111, USA}
\author{{\L}ukasz Cywi{\'n}ski}
\affiliation{Condensed Matter Theory Center, Department of Physics, University of Maryland, College Park, Maryland 20742-4111, USA}
\affiliation{Institute of Physics, Polish Academy of Sciences, al.~Lotnik{\'o}w 32/46, PL 02-668 Warszawa, Poland}
\author{Dimitrie Culcer}
\affiliation{Condensed Matter Theory Center, Department of Physics, University of Maryland, College Park, Maryland 20742-4111, USA}
\author{Xuedong Hu}
\affiliation{Condensed Matter Theory Center, Department of Physics, University of Maryland, College Park, Maryland 20742-4111, USA}
\affiliation{Joint Quantum Institute, Department of Physics, University of Maryland, College Park, Maryland 20742-4111, USA}
\affiliation{Department of Physics, University at Buffalo, SUNY, Buffalo, New York 14260-1500, USA}
\author{S. Das Sarma}
\affiliation{Condensed Matter Theory Center, Department of Physics, University of Maryland, College Park, Maryland 20742-4111, USA}
\affiliation{Joint Quantum Institute, Department of Physics, University of Maryland, College Park, Maryland 20742-4111, USA}
\begin{abstract}
We study exchange coupling in Si double quantum dots, which have been proposed as suitable candidates for spin qubits due to their long
spin coherence times. We discuss in detail two alternative schemes which have been proposed for implementing spin qubits in quantum dots. One
scheme uses spin states in a single dot, and the interdot exchange coupling controls interactions between unbiased dots. The other scheme
employs the singlet and triplet states of a biased double dot as the two-level system making up the qubit and exchange controls the energy
splitting of the levels. Exchange in these two configurations depends differently on system parameters. Our work relies on the Heitler-London
approximation and the Hund-Mulliken molecular orbital method. The results we obtain enable us to investigate the
sensitivity of the system to background charge fluctuations and determine the conditions under which optimal spots, at which the influence of
the charge noise is minimized,  may exist in Si double quantum dot structures.

\end{abstract}
\maketitle

\section{Introduction}

The use of quantum-mechanical two-level systems to construct quantum bits, or qubits, for the storage and manipulation of information has
attracted considerable attention in the past decade. Computational algorithms relying on the principles of quantum mechanics such as unitary
evolution, entanglement, and coherent superposition are expected to offer novel solutions to problems such as factoring, searching, and
simulating quantum-mechanical systems.\cite{Nielsen_Chuang}  Among the many physical systems that have been proposed as candidates for qubits,
solid state systems are particularly promising because of their potential for scalability, and spin-based qubits in solid state systems are
regarded as ideal candidates due to their observed long coherence times.\cite{Petta_Science05,Koppens_PRL08}

A common essential feature of the most prominent spin-based quantum computing (QC) architectures is the employment of exchange interaction
between localized electrons. For example, in the original Loss-DiVincenzo (LDV) proposal\cite{Loss_PRA98} exchange interaction is pulsed on and
off by electrical control and produces two-qubit gates such as SWAP and controlled-NOT gates. The Kane proposal\cite{Kane_Nature98} of using
nuclear spins of $^{31}$P in Si as qubits employs donor electron exchange to mediate nuclear spin interactions and produce two-qubit gates.  In
the more recent singlet-triplet (ST) qubit proposal,\cite{Taylor_NP05,Petta_Science05,Taylor_PRB07,Hanson_RMP07} controlled exchange splitting
between singlet and triplet states leads to single-qubit gates.  Exchange interaction is electrostatic in nature so that it can be strong and
can produce fast gates (on a time scale as short as tens of picoseconds), and it can be controlled electrically by changing voltages applied to
the gates defining the quantum dots.  In other words, exchange gates are easy to control and allow straightforward interface with existing
microelectronic devices. A potentially important shortcoming of an exchange-based QC architecture is also related to its electrostatic
character: turning on the exchange interaction could potentially make the system vulnerable to electrical fluctuations in the environment
(background charge noise,\cite{Burkard_PRB99,Coish_PRB05,Hu_PRL06} gate noise,\cite{Hu_PRA00} electron-phonon
interaction,\cite{Roszak_Preprint09,Hu_Preprint09} etc.), which can lead to qubit decoherence. Therefore, it is imperative that the exchange
interaction be characterized carefully, so that its magnitude is known in a physical system, and possible ranges of control parameters for which
the influence of the external charge noise is minimized (so-called optimal points) can be identified.\cite{Coish_PRB05,Hu_PRL06,Stopa_NL08}

One of the most promising semiconducting materials to host a spin quantum-information processor is silicon,\cite{DasSarma_SSC05} which has
outstanding spin coherence properties in the bulk.\cite{Tyryshkin_JPC06}  Natural silicon contains only a small number of nuclear spins (5\% of
$^{29}$Si, which has spin $1/2$), and can be further isotopically purified.  The strength of the hyperfine interaction between a conduction
electron (or a donor-bound electron) with each individual nucleus is also weak.\cite{Asalli_preprint}  As such the hyperfine-interaction induced
electron-spin decoherence, dominant in III-V materials such as GaAs,\cite{Khaetskii_PRB03, Witzel_PRB06, Yao_PRB06, Zhang_PRB06,Deng_PRB06,
Deng_PRB08, Coish_PRB08, Cywinski_PRL09, Cywinski_PRB09} is suppressed in Si.  Furthermore, Si has a small spin-orbit
interaction\cite{Tahan_PRB05} and no piezoelectric interaction, which together means slower electron spin relaxation due to
phonons.\cite{Prada_PRB08}  Silicon does have a drawback in its somewhat complicated conduction band structure with six equivalent minima. This
orbital degeneracy could lead to difficulties such as the atomic-scale position dependence of exchange interaction between donor electrons.\cite{Koiller_PRL02}
At present there are several proposed Si quantum-computer architectures, including architectures based on donor electron or nuclear spins in a
Si:P system,\cite{Kane_Nature98, Vrijen_PRA00,Morello_PRB09} single electron spins in gate-defined quantum dots in Si/SiGe\cite{Friesen_PRB04}
or Si/SiO$_2$(Ref. 33) quantum dots,\cite{Liu_PRB08,Liu_APL08,Nordberg_MOS_09} and SiGe nanowires.\cite{MarcusGroup_NatureNano07}

The development of Si single-electron devices suitable for quantum-information processing is a relatively recent phenomenon because of the
difficulty in identifying and growing appropriately lattice-matched barrier materials (as compared to Ga$_{1-x}$Al$_x$As for GaAs), which help
minimize carrier scattering near the interface of a heterojunction, and difficulties in fabricating and characterizing single-donor devices at
the atomic length scale.  Nevertheless, exciting experimental progress has been made in both donor-based devices\cite{Andresen_NanoLett07,
Mitic_Nanotech08, Kuljanishvili_NP08, Lansbergen_NP08, Fuhrer_NanoLett09,Tan_preprint09} and gate-defined quantum dots.\cite{Goswami_NP07,
Simmons_APL07, Shaji_NP08, Liu_PRB08, Liu_APL08, Nordberg_MOS_09, Lim_APL09}

Motivated by the experimental attempts to realize spin qubits in a Si metal-oxide-semiconductor field-effect transistor (MOSFET),\cite{Liu_PRB08, Liu_APL08, Nordberg_MOS_09} here we consider the
prospects for creating spin qubits in a lateral double quantum dot (DQD) structure fabricated in a silicon MOSFET by lithographic patterning. We
calculate the exchange coupling and tunnel coupling of electrons in DQD structures in a Si/SiO$_2$ system under the assumption of a reasonably
large valley splitting.  We first analyze the applicability of the Heitler-London (HL) and Hund-Mulliken (HM) models for calculating exchange
splitting in a Si double dot, then perform our calculations for experimentally achievable sizes of dots. We study two types of DQD
configurations. One is a symmetric DQD, where the single-electron levels of the two dots are on resonance and are appropriate for the exchange
gates of the original LDV proposal. The other type is a biased DQD, where the two dots are voltage-biased to the vicinity of the resonance
between the two-dot singlet and one of the doubly occupied singlet states.  This is a configuration appropriate for the ST qubit architecture.

Compared to the calculation of exchange coupling in GaAs,\cite{Burkard_PRB99,Hu_PRA00,Scarola_PRA05,Wiel_NJP06,Pedersen_PRB07,Stopa_NL08} there
are some distinctive features for such a calculation in Si. First of all, in a Si MOSFET structure the Coulomb interaction is enhanced by the
proximity of a lower-$\kappa$ dielectric, while the kinetic energy is reduced by the larger effective mass.  Therefore correlation effects
should be much stronger in a Si double dot as compared to a GaAs structure, shrinking the range of validity for the simpler models of exchange
calculation (as we will discuss in Sec.~\ref{sec:reliability}).  Indeed, identifying these ranges is one of the primary aims of the present
calculations. Another important distinction is the presence of conduction-band valleys in Si. Bulk Si has six degenerate conduction-band minima.
Confinement and uniaxial strain in the $\hat{\bm z}$ direction split these six valleys into a two-dimensional(2D) manifold of lower energy and a
four-dimensional manifold of higher energy that is separated from the lower one by several tens of meV(Ref. 31 and 51) and is
typically neglected in discussions of the lowest-energy states.  Furthermore, the interface potential typically splits the two lower-energy
valleys by the energy on the order of a fraction of an meV.\cite{Ando_RMP82, Friesen_PRB07, Saraiva_PRB09} The presence of the valley degree of freedom
complicates the energy spectrum and spin structure of the two-electron states in silicon quantum dots,\cite{Hada_PRB03,Hada_JJAP04,Culcer_09}
and can potentially affect the effectiveness of Si as a host for spin-based quantum-information processing.

In this paper we consider situations in which the valley splitting $\Delta$ (between the ground and the first valley eigenstates) is reasonably
large so that consistent loading of the nondegenerate ground valley-eigenstate in a quantum dot or a double dot can be achieved.  We also
confine ourselves to the situation where the valley composition of the ground valley eigenstate is uniform throughout the DQD.\cite{Culcer_09}
Under these conditions we can perform our calculation of exchange within a single valley because the two-electron Hamiltonian Eq.~(\ref{eq:H})
does not have any intervalley matrix elements: all the single-particle terms are slowly varying in space, and all the Coulomb matrix elements
(except one exchange integral) vanish exponentially $\sim \exp(-L^2/a_{0}^2)$ where $L$ is the $\hat{\bm z}$-direction confinement and is on the order of
3 nm while $a_{0}$ is the Si lattice constant of 5.43{\AA}, so that the exponential factor is extremely small.  The lone exchange integral
that does not vanish exponentially is very small in any case ($<$ 0.2$\mu$eV for reasonable QD sizes) so that we take it as zero as well. Since
the double-dot Hamiltonian does not lead to any coupling to the higher-energy valley states even if an excited valley is nearby, it will not
produce any modification in the single-valley exchange we calculate, and the basic physics we consider here is the same as in a single-valley
system such as GaAs.

In this study we also investigate the sensitivity of exchange-coupled spin qubits to environmental electrical fluctuations.  As we mentioned
above, finite-exchange coupling makes spin qubits susceptible to electrical noises from sources such as gate electrodes and background charge
fluctuations.  Background charge fluctuations arise due to trapped electrons in defect sites at semiconductor interfaces.  Movement of nearby
trapped charges affects quantum dot systems in two ways: rise or fall of the barrier between the dots and fluctuations in the two confining
potentials of the dots. Due to the possible importance of charge noise (SiO$_{2}$, the most commonly employed barrier material for silicon
heterojunctions, is a well-known source of charge noise), the conditions for the existence of the so-called \textit{optimal points} (in analogy
to the optimal point studied for superconducting charge qubits, where charge-noise-induced dephasing is minimized\cite{Vion_Science02}), where
the exchange splitting for the two-electron states is the least sensitive to environmental charge fluctuations, need to be investigated. In this
work we therefore also calculate the sensitivity of the qubit to the charge noise and identify the optimal points in the qubit parameter space
in which the qubit is less sensitive to fluctuations in gate voltages.\cite{Coish_PRB05,Hu_PRL06,Stopa_NL08,Ramon_09}

The goal of this paper is a qualitative (and at best semi-quantitative) description of the exchange-gate behavior in Si DQD qubit architecture
as a function of the relevant physical parameters: dot separation, confinement potential, dot size, interdot bias, and external magnetic field.
Since the details of the actual Si DQD confinement in real samples are unknown and likely to be extremely complicated, we use a simple
physically motivated model potential to study the problem.  Given the simplicity of our model, our calculation of exchange also employs
well-known approximation schemes (i.e. Heitler-London and Hund-Mulliken) which should be qualitatively valid in various limits.  At this stage
of the development of Si spin qubits, it is not particularly meaningful to carry out a fully numerical Schr\"{o}dinger-Poisson-configuration-interaction modeling of the exchange-gate energetics, because the details of the structures are simply not known and will probably remain
unknown for some time at the desired level of accuracy to justify a completely computational first principles approach.  Our simpler approach
with relative computational ease enables us to provide a thorough exploration of parameter dependence of the Si DQD exchange-gate energetics,
which should be helpful in this early stage of experimental development. In this sense, our work should be considered as the Si equivalent to
the corresponding GaAs qubit theories carried out in Refs.~\onlinecite{Burkard_PRB99} and \onlinecite{Hu_PRA00}.

The outline of this paper is as follows. Section II introduces the formalism that will be used throughout the paper, which includes the model
potentials for double quantum dots, the procedures followed in the determination of exchange coupling, and the discussion of their limits of
validity. In Sec. III the two architectures employed to make spin qubits in quantum dots are discussed. Section IV contains our results on
exchange coupling and sensitivity to noise. Section V contains a summary and conclusions.

\section{Theoretical formalism }\label{sec:formalism}
\subsection{Hamiltonian and double-dot-model potentials}
Consider a DQD at the Si/SiO$_2$ interface with growth direction along the $\hat{\bm z}$-axis. The effective-mass Hamiltonian for two electrons in a DQD is then
\begin{equation}
\hat{H}=\sum_{i=1,2} \hat{h}_i +\dfrac{e^2}{\kappa r_{12}}  = \sum_{i=1,2} \hat{h}_i + \hat{C}
\label{eq:H}
\end{equation}
where $i \! =\! 1,2$ labels the two electrons, $\hat{h}_{i}$ is the single-particle Hamiltonian, $\kappa = (\varepsilon_{Si} +
\varepsilon_{SiO_2})/2$ is the effective dielectric constant including the image charge in the adjacent SiO$_{2}$, and $r_{12}$ is the distance
between the two electrons.  The single-electron Hamiltonian is
\begin{eqnarray}
\hat{h}_i & = & \hat{T}_{i} +V({\bf{r}}_i) + eEx_{i} + g_{\text{eff}}\mu_{\text{B}}B S_{iz} \,\, ,  \\
\hat{T}_{i} & = & \frac{1}{2m} [{\bf p}_i-\frac{e}{c}{\bf A}({\bf r}_i)]^2 \,\, ,
\end{eqnarray}
where $m\! = \! 0.191 m_e$ is the transverse (isotropic in the $xy$ plane) effective mass for electrons in the $\hat{\bm z}$ and $-\hat{\bm z}$
valleys (the valleys contributing to the lowest-energy valley eigenstate) in Si, ${\bf A}=B(-y,x,0)/2$ is the vector potential for magnetic
field $B$ applied along the $\hat{\bm z}$-direction, $V({\bf r}_i)$ is the confinement potential chosen to approximate the real potential for
two dots separated by a potential barrier, $E$ is the electric field applied in the $\hat{\bm x}$-direction (the interdot axis direction of the
DQD), and the last term is the electron Zeeman energy (here we have assumed that both electrons are experiencing the same magnetic field), where
for Si we can take $g_{\text{eff}} \! =\! 2$.

In the above $\hat{h}_{i}$ we have omitted the spin-orbit coupling terms\cite{Stepanenko_PRB03} since these are expected to be very small in
silicon\cite{Wilamowski_PRB02,Tahan_PRB05} compared to GaAs.  The spin-orbit terms in general induce an anisotropic (Dzyaloshinskii-Moriya-type) exchange interaction between the two spins,\cite{Kavokin_PRB01,Gorkov_DM_PRB03,Stepanenko_PRB03,Kunikeev_PRB08,Baruffa_09} which could
affect the performance of two-qubit exchange gates.\cite{Chutia_PRB06} While an experiment aiming at detecting this kind of anisotropic exchange
interaction was proposed,\cite{Chutia_PRB06} these effects have not yet been clearly shown to be relevant to experiments on GaAs (where they are
expected to be much stronger than in Si). Thus, at this stage of research on Si DQDs we choose to neglect them in this work.

The potential $V(\bf{r}_i)$ describes the DQD system of two dots located in the $xy$-plane which are tightly confined in the $\hat{\bm z}$ direction.  The
exact form of the confinement along the $\hat{\bm z}$ direction is irrelevant here as long as the characteristic localization length in the
$\hat{\bm z}$-direction is much smaller than in the $\hat{\bm x}$ and $\hat{\bm y}$ directions.  This is fulfilled in Si/SiO$_{2}$ DQD where
the $z$-confinement length is on the order of 3 nm, while the confinement length in the $\hat{\bm x}$ and $\hat{\bm y}$-directions is at least
several times larger.  The strain at the Si/SiO$_2$ interface and the confinement in the $\hat{\bm z}$-direction lifts the sixfold valley
degeneracy in bulk silicon and leaves the ground valleys doubly degenerate.

This ground-valley degeneracy is further split by an amount $\Delta$ due to the interface potential,\cite{Ando_RMP82,Friesen_PRB04,
Srinivasan_APL08, Friesen_09,Saraiva_PRB09,Culcer_09} which allows us to focus only on the lowest-energy valley eigenstate. At this stage we take the 2D approximation in which the finite confinement length in the $\hat{\bm z}$-direction is
neglected.\cite{Burkard_PRB99}  The reliability of this approximation will be ascertained in Sec.~\ref{sec:reliability}.  We will use three
different models of the DQD potential.  The first is a quartic model\cite{Burkard_PRB99}
\begin{equation}
 V({\bf r})= \frac{1}{2} m \omega_0^2 \bigg[\frac{1}{4d^2}(x^2-d^2)^2+ y^2\bigg] \,\, ,
\label{eq:quartic}
\end{equation}
while the second is a biquadratic model\cite{Pedersen_PRB07}
\begin{equation}
V({\bf r})= \frac{1}{2}m \omega_0^2
\left(\text{Min}\left[\left\{(x-d)^2,(x+d)^2\right\}\right]+y^2\right) \,\,.
\label{eq:biquad}
\end{equation}
Note that both of these potentials provide infinite confinement (i.e.~we have $V(r) \! \rightarrow \! \infty$ as $r$ goes to infinity).  A
third, more realistic potential, will be introduced in Sec.~\ref{sec:reliability}, where we will use the comparison of calculations with all
three model potentials to gauge the limits of validity of our theoretical approach.

For the quartic potential from Eq.~(\ref{eq:quartic}) the two dots are separated by a distance $2d$ and the central barrier height is
$V_b=\frac{1}{8} m \omega_0^2 d^2$. For the biquadratic potential from Eq.~(\ref{eq:biquad}) the distance between the dots is the same, while
the central barrier height is $V_b=\frac{1}{2} m \omega_0^2 d^2$, four times larger than for the quartic potential.  Note that for these
potentials the barrier height and the interdot distance are not independent variables. On the other hand, in experiments there are typically
three independent voltages controlling the DQD potential (two gates controlling the depth of the left and right dot potentials, and an
additional gate separately controlling the barrier). However, most often only two variables (the left-/right-dot gate voltage) are changed
simultaneously, which we model with a nonzero in-plane electric field $E$ in Eq.~(\ref{eq:H}).

Each of the model potentials is designed in such a way that around the centers of the two dots at $x \! = \! \pm d$ (for $E \! = \! 0$) the
confinement is approximately parabolic so that the ground state single-particle wave-functions are harmonic oscillator states in the isolated
dot limit.  More precisely, near the centers of the two dots we have
\beq
V(x\approx \pm d) \approx \frac{1}{2}m\omega^{2}_{0} [ (x\mp d)^{2} + y^{2} ] \equiv V_{R/L}(\mathbf{r})  \,\, ,
\eeq
and the approximate single-dot lowest-energy eigenstates $ \varphi_{L/R}(\mathbf{r})$ are solutions of the equation
\beq
[ \hat{T}_{i} + V_{L/R} (\mathbf{r}) ] \varphi_{L/R}(\mathbf{r}) = \hbar\omega_{0}  \varphi_{L/R}(\mathbf{r})
\eeq
with $\varphi_{L/R}(\mathbf{r})\! = \! \left< \mathbf{r}| L/R\right> $ being the  Fock-Darwin states centered on the right/left
dot\cite{Burkard_PRB99}
\beq \varphi_{L/R} = \frac{1}{a}\sqrt{\frac{b}{\pi}}\exp[\mp \frac{iy d\sqrt{(b^2-1)}}{a^{2}}]\exp[-\frac{b((x\pm
d)^2 +y^2)}{2a^{2}}]  \label{eq:phi} \eeq
where the first exponential is the magnetic phase for a displaced orbital, $a \! \equiv \! \sqrt{\hbar/m\omega_{0}}$ is the Fock-Darwin radius,
and $b \! \equiv \! \sqrt{1+\omega^{2}_{L}/\omega^{2}_{0} }$ is a measure of the degree of magnetic confinement, where $\omega_{L} \! = \!
eB/2mc$ is the electron Larmor frequency. The overlap of the right and left single-particle orbitals is
\beq
l \equiv \langle L | R \rangle = \exp\left[ \frac{d^{2}}{a^{2}}\bigg(\frac{1}{b} - 2b\bigg)\right] \,\, .
\label{eq:overlap}
\eeq
For example, at zero magnetic field the overlap is $l=\exp(-d^2/a^2)$.

\begin{figure}
\includegraphics[width=0.99\linewidth]{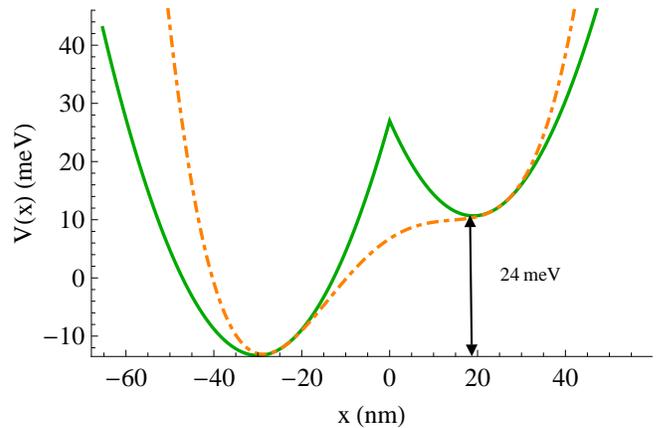}
\caption{ (color online) The potential of an asymmetric double quantum dot in the $\hat{\bm x}$-direction ($\hbar\omega_{0}=6$ meV). The solid line corresponds to the
biquadratic potential while the dot-dashed line corresponds to the quartic potential. The ratio of the detuning to the single-particle energy
near the anti-crossing (explained in the text) is $\epsilon/\hbar\omega_{0} \approx 4$.} \label{fig:asymmetric}
\end{figure}

A ST qubit is operated in the parameter regime where the lowest-energy (1,1) singlet (one electron in each dot) is close in
energy to the lowest energy (2,0) singlet state (both electrons in the left dot, as shown in Fig.~\ref{fig:asymmetric}).  For small quantum dots
the on-site Coulomb repulsion is quite strong, so that the DQD needs to be strongly biased and asymmetric.  We model such a biased DQD by adding
an in-plane electric field ${\bm E} = E\hat{\bm x}$ along the inter-dot axis to the Hamiltonian. While adding $E$ displaces the two minima of
the potential corresponding to the centers of the two dots for both model potentials, other responses are significantly different for these
models. Specifically, when an electric field is added to the quartic potential, the potential minima of $V(x)$ are altered by $E$ and thus the curvatures of $V(x)$ near the potential minima are also changed, which in turn changes the single-dot confinement energies of the L/R states (for example, in
Fig.~\ref{fig:asymmetric} we show a case in which there is no minimum on the right side, only a saddle point). In addition, as shown in
Fig.~\ref{fig:asymmetric}, the height of the barrier between the two dots is strongly affected by $E$ in the quartic model.  In short the
quartic model of the DQD becomes rather cumbersome when the two dots are biased. These observations lead us to adopt the biquadratic model for
the calculations with finite bias field $E$.  It is reasonable to expect that the behavior of this model is closer to the experimental
situation, in which the primary effect of changing the gate voltages is to introduce an offset between the ground state energies of the two
dots.  Changes in the barrier height (i.e., ~the tunnel coupling between the dots) and in the confinement energy in each dot are secondary
effects.

Using the biquadratic potential from Eq.~(\ref{eq:biquad}) with an added electric field $E$, the two minima of the potential (corresponding to
the centers of the two dots) are shifted by the same distance $f a$, with $f$ given by
\beq
f \equiv \frac{eEa}{\hbar \omega_{0}} \,\, .
\eeq
The single-dot orbitals are now given by Eq.~(\ref{eq:phi}) with $d$ replaced by $d - fa$ for the right dot, and $-d$ replaced by $-d - f a$ for
the left dot.

In experiments on asymmetric DQDs the commonly used parameter is the so-called detuning energy $\epsilon$.  It parameterizes the changes of the
DQD potential along a specific line in the experimentally obtained charge-stability diagram\cite{Hanson_RMP07} of the DQD.  A change in
$\epsilon$ corresponds to a change in the difference between the single-particle ground states in the two dots.  Using the biquadratic potential model, we get $\epsilon  \!
\equiv \! 2 f \hbar \omega_0 (d/a) \! = \! 2eEd$.

\subsection{Theoretical approaches for calculating the exchange splitting}
\label{sec:HL_HM}

In a uniform magnetic field, if two electrons interact only through the Coulomb interaction, the spin eigenstates are the singlet ($|S\rangle =
\frac{1}{\sqrt{2}} | \uparrow \downarrow - \downarrow \uparrow \rangle$, with total spin $S=0$) and triplet ($|T_{0,+,-}\rangle =
\frac{1}{\sqrt{2}} |\uparrow \downarrow + \downarrow \uparrow \rangle$, $| \uparrow \uparrow \rangle$, and $| \downarrow \downarrow \rangle $),
which have total spin $S=1$) states. The corresponding spatial parts of the wave functions are symmetric and antisymmetric under exchange of the
two electrons, respectively. For quantum computing, where orbital degrees of freedom are frozen out, the energy splitting $J \equiv E_{T} -
E_{S}$ between the unpolarized triplet state $|T_{0}\rangle$ and the ground singlet state is the key physical quantity and we will refer to it as the
exchange splitting or, in general, exchange interaction.

To calculate the exchange splitting $J$, the two-electron Hamiltonian should be diagonalized in the singlet and triplet subspaces, respectively
(with symmetric and antisymmetric orbital-state bases), which amounts to a full configuration interaction (CI) calculation.  Since there is an
infinite number of two-electron states, all practical CI calculations are truncated. Depending on the degree of truncation, there is a hierarchy
of models for the calculation of exchange interaction in a two-electron double quantum dot.

Interestingly, the reliability of a CI calculation is not completely determined by the size of the finite basis left after the
truncation.\cite{Herring_Magnetism}

One particular test of validity of the given approach to calculating $J$ is the Lieb-Mattis theorem,\cite{Mattis} which dictates that at zero
magnetic field the singlet must always be the ground state so that $J > 0$. The condition for applicability of this theorem in 2D and three-dimensional (3D) cases is for
the confinement potential to be separately symmetric, i.e., ~symmetric under exchange of each of the $x$, $y$, and $z$ coordinates of the two
particles. This is the case here, since we use model confinement potentials which are separable. At finite $B$ either sign of $J$ is
possible.\cite{Burkard_PRB99,Hu_PRA00,Scarola_PRA05,Wiel_NJP06}

\subsubsection{Heitler-London approximation}

The simplest approach for calculating the exchange splitting $J$ in a DQD is the HL method,\cite{Mattis,
Burkard_PRB99,Hu_PRA00,Calderon_PRB06} where the ground two-electron double-dot singlet and triplet states are built from single-dot
single-electron states in the simplest possible combination. The spatial parts of the normalized singlet and triplet states are then given by
\beq
\ket{S/T}= \dfrac{ \ket{L(1)R(2)}  \pm  \ket{L(2)R(1)} }{\sqrt{2(1 \pm l^2})} \,\, ,
\label{eq:phi_HL}
\eeq
where $\ket{L/R(i)}$ are the single-dot ground state orbitals for the $i$th electron, given by Eq.~(\ref{eq:phi}) for the symmetric DQD case,
and $l \! \equiv\! \langle L | R \rangle$ is  the overlap between the left and right orbitals given in Eq.~(\ref{eq:overlap}). Heitler-London
wave functions are essentially educated guesses of the two-electron ground states. They are reasonable wave functions for quantum dot states at
large interdot separations.

In the HL basis the exchange splitting is the energy difference between the unpolarized $T_{0}$ triplet and the singlet states
\begin{equation}
 J_{\text{HL}}= \langle T_{0} |\hat{H}|T_{0}\rangle-\langle
S|\hat{H}| S \rangle \,\, ,
\end{equation}
which can be written as
\beq
J_{\text{HL}} = \frac{2l^{2}}{1-l^{4}} \left( W_{v} + D_{0} - \frac{1}{l^2}E_{0}   \right) \,\, ,
\label{eq:J_HL}
\eeq
where $W_{v}$ is the single-particle contribution due to the confinement potential
\beq W_{v} = \left\langle L(1)R(2) \Big| \hat{v} \Big| L(1)R(2) -
\frac{1}{l^2}L(2)R(1) \right\rangle \,\, , \label{eq:Wv} \eeq
with $v(1,2) \! \equiv \! V(1) + V(2) -V_{R}(2)-V_{L}(1)$. The analytical expressions for $W_{v}$ for quartic and biquadratic potentials are
given in Appendix.  Physically, $W_v$ represents the gain in kinetic energy for the singlet state arising from the fact that
electrons in this state are spread out more evenly across the two dots.  $D_{0}$ is the direct Coulomb interaction contribution
\beq
D_{0} =  \langle L(1)R(2)| \hat{C} | L(1)R(2) \rangle \,\, ,
\label{eq:D0_def}
\eeq
and $E_{0}$ is the exchange integral
\beq
E_{0} = \langle L(1)R(2)| \hat{C} | L(2)R(1) \rangle \,.
\label{eq:E0_def}
\eeq
Both Coulomb terms are positive.  Together they favor the triplet state since in an antisymmetric state the two electrons avoid each other to
minimize their interaction.  Both Coulomb integrals $D_{0}$ and $E_{0}$ can be done analytically in 2D (Ref. 9) using the states
given in Eq.~(\ref{eq:phi}).  In units of the effective Rydberg $\text{Ry}^{*} \! \equiv \! m e^{4}/2\hbar^{2}\kappa^{2}$ (approximately $44.76$
meV in a Si/SiO$_{2}$ system) we obtain
\begin{eqnarray}
\tilde{D}_{0} & = & \frac{\sqrt{2\pi b}}{\tilde{a}} I_{0} \left( \frac{b\tilde{d}^{2}}{\tilde{a}^{2}} \right) e^{-b \tilde{d}^{2}/\tilde{a}^{2}}
\,\, ,\label{eq:D0}  \\
\tilde{E}_{0} & = &  l^2 \frac{\sqrt{2\pi b}}{\tilde{a}}\, e^{\frac{\tilde{d}^2}{\tilde{a}^2}(b-\frac{1}{b})} I_{0}
\bigg[\frac{\tilde{d}^2}{\tilde{a}^2}\bigg(b-\frac{1}{b}\bigg)\bigg] \,\, ,
\label{eq:E0}
\end{eqnarray}
where $I_{0}$ is the zeroth-order modified Bessel function.  $\tilde{a}\! \equiv \! a/a^{*}_{\text{B}}$ and $\tilde{d}\! \equiv \! \
d/a^{*}_{\text{B}}$ are the Fock-Darwin radius and the half interdot distance expressed in units of the effective Bohr radius $a^{*}_{\text{B}}
\! \equiv \! \hbar^{2}\kappa/me^{2}$ (approximately $2.11$ nm in a Si/SiO$_{2}$ system).  Since $\tilde{a}/\tilde{d} = a/d$, we can drop the
tildes when the ratio appears in Eqs.~(\ref{eq:D0}) and (\ref{eq:E0}).  The contributions of $D_{0}$ and $E_{0}$ to $J$ can be of comparable
magnitude. For example, at large distances ($d\! \gg \! a$), we have $\tilde{E}_{0} \! \sim \! l^{2}\tilde{D}_{0} \tilde{d}/\tilde{a}$ at $B\! =
\!0$, and $\tilde{E}_{0} \! \approx \! l \tilde{D}_{0}$ at high $B$ fields (when $b\! \gg \! 1$).  The $1/l^2$ factor in front of $E_{0}$ in
Eq.~(\ref{eq:J_HL}) makes the negative contribution of the exchange integral $E_{0}$ significant, and possibly larger in absolute magnitude than
the direct term $D_{0}$ (this clearly occurs at high $B$, when $J$ can become negative).

\subsubsection{Hund-Mulliken Approximation}

The HL approximation neglects both the contribution of the higher-energy orbitals from the two dots and the possibility of double electron
occupation on either of the dots. The latter restriction is removed in the Hund-Mulliken (HM)
method,\cite{Mattis,Burkard_PRB99,Hu_PRA00,Wiel_NJP06,Pedersen_PRB07,Hatano_PRB08} in which the doubly-occupied states (built out of the
ground-state single-dot orbitals) are included in the basis set.  This is of particular importance when we consider the asymmetric case of a
biased DQD, which is relevant for the experiments on singlet-triplet qubits. In that system the mixing between the delocalized singlet $S(1,1)$
and one of the doubly-occupied states [e.g.~$S(2,0)$] by the tunneling of an electron between the two dots is the main mechanism that brings
about the singlet-triplet splitting $J$ [$(n_{L}$,$n_{R})$ denotes the number of electrons in each dot].

In the HM method we take into account the doubly-occupied $(2,0)$ and $(0,2)$ states.  We still employ only the lowest-energy single-dot
orbitals, so that the doubly-occupied states must be spin singlets (their spatial parts being symmetric). These two doubly-occupied states,
together with the $(1,1)$ spin-singlet state, form the basis of three singlet states in the HM method, which have an anti-symmetric spin part
but a symmetric spatial wave function.  Conversely, the spin-triplet state, with a symmetric spin part, has only one possible spatial
configuration since the electrons cannot occupy any higher-energy orbital states. It is convenient to use the basis of orthonormalized
single-dot states, which we define following Ref.~[\onlinecite{Burkard_PRB99}] as $\Phi_{L/R}=(\varphi _{L/R}-g \varphi_{R/L})/\sqrt{1-2 lg+
g^2}$, where $g=(1-\sqrt{1-l^2})/l$. Then the spatial wave functions of the three singlets and the $(1,1)$ triplet are given by
\begin{eqnarray}
\Psi_{L/R}^{d}(\mathbf{r}_1,\mathbf{r}_2) & = &\Phi_{L/R}(\mathbf{r}_1)\Phi_{L/R}(\mathbf{r}_2) \\
\Psi^{(1,1)}_{S/T}(\mathbf{r}_1,\mathbf{r}_2) & = &\frac{1}{\sqrt{2}} \Big[\Phi_{L}(\mathbf{r}_1)\Phi_{R}(\mathbf{r}_2) \pm
\Phi_{R}(\mathbf{r}_1)\Phi_{L}(\mathbf{r}_2) \Big] ~,~~~
\end{eqnarray}
where the superscript $d$ denotes the doubly occupied states.  The Hamiltonian written in the $S(2,0)$, $S(0,2)$, $S(1,1)$, $T(1,1)$ basis takes
the form
\begin{equation}
H=\varepsilon_R+\varepsilon_L+ \left(
\begin{array}{llll}
U-\varepsilon&X&\sqrt{2}t&0\\
X&U+\varepsilon&\sqrt{2}t&0\\
\sqrt{2}t&\sqrt{2}t&V_{S}&0\\
0&0&0&V_{T}
\end{array}
\right)
\label{eq:H_HM}
\end{equation}
where
\begin{eqnarray}
\begin{array}{c l l l c}
\epsilon = \varepsilon_R-\varepsilon_L
\\
\\
\varepsilon_R=\langle \Phi_{R}|\hat{h}|\Phi_{R}\rangle \ \ \ \ \    \varepsilon_L=\langle \Phi_{L}|\hat{h}|\Phi_{L}\rangle
\\
\\
X = \langle \Psi_{L/R}^d|\hat{C}|\Psi_{R/L}^d\rangle  \ \ \ \ \  U = \langle \Psi_{L/R}^d|\hat{C}|\Psi_{L/R}^d\rangle
\\
\\
V_{S} = \langle \Psi_S |\hat{C}| \Psi_S \rangle  \ \ \ \ \ \ \  V_{T} = \langle \Psi_T |\hat{C}| \Psi_T \rangle
\\
\\
t = \langle \Phi_{L/R}|\hat{h}|\Phi_{R/L}\rangle +  \frac{1}{\sqrt{2}} \langle \Psi_S |\hat{C}| \Psi_{L/R}^d \rangle = t' + w
\end{array}
\label{eq:matrix}
\end{eqnarray}
In the above, $\varepsilon_{L}$ and $\varepsilon_{R}$ are the single-particle energies in the left and right dots, $\epsilon$ is the detuning
parameter, $X$ is similar to a interdot Coulomb exchange integral, $U$ is the on-site (Hubbard) Coulomb repulsion, $V_{S}$ and $V_{T}$ are the
Coulomb energies for the (1,1) singlet and triplet states, with $\Psi^{(1,1)}_{S/T}$ built out of orthogonalized single-electron orbitals, and
$t$ is the interdot tunneling matrix element modified by the Coulomb matrix element $w$ ($t'$ is the bare tunneling).  All these can be
expressed in terms of the analogous matrix elements between the original, nonorthogonal, left and right orbitals $\varphi_{L/R}$ (these
bare-orbital matrix elements are termed $D_{0}$, $E_{0}$, $U_{0}$ etc). Explicit formulas are given in Appendix.

Diagonalizing the Hamiltonian given in Eq.~(\ref{eq:H_HM}) can then give the exchange interaction which is the energy difference between the
two lowest eigenvalues [one corresponds to a singlet, the other the (1,1) triplet].  For the symmetric case in zero $B$ field, the lowest
eigenvalue $\varepsilon_{S_-}$ is dominated by the energy of the $S(1,1)$ spin singlet state while modified by the two doubly occupied states.
On the other hand, for the asymmetric (biased) DQD, $\varepsilon_{S_-}$ is dominated by a mixture of the $S(2,0)$ and $S(1,1)$ singlet states
[we assumed here that the bias is such that $S(2,0)$ state is the lower-energy doubly occupied state, i.e., ~the potential of the left dot is
lowered to attract an additional electron].  $S(0,2)$ state can then be dropped from our consideration.

We define the critical bias detuning $\epsilon_c$ as the bias at which the $S(1,1)$ state is on resonance with the doubly occupied $S(2,0)$
state when tunneling $t$ is neglected. Then from Eq.~(\ref{eq:H_HM}), the energy of $S(1,1)$ is
\begin{equation}
\varepsilon_{S(1,1)} \simeq \varepsilon_R + \varepsilon_L + V_S \,\, ,
\end{equation}
and the energy of the $S(2,0)$ state is
\begin{equation}
\varepsilon_{S(2,0)} \simeq 2\varepsilon_L+U \,\, .
\end{equation}
Therefore the critical bias detuning $\epsilon_c$ corresponds to the energy difference $\epsilon_c \! = \! U-V_S$, at which
$\varepsilon_{S(1,1)}=\varepsilon_{S(2,0)}$.  When $\epsilon \ll \epsilon_c$, the ground state is composed mostly of $S(1,1)$; when $\epsilon
\gg \epsilon_c$ the $S(2,0)$ state becomes the ground state.  Around the anticrossing region where $\epsilon \! \approx \! \epsilon_{c}$ the
$(1,1)$ singlet and $(2,0)$ singlet states are strongly mixed.

We note here that the bias regime in which we are interested in our HM calculations is where the S(1,1) singlet is the dominant component of the
ground singlet state.  This is the regime where single-qubit gates are performed for an ST qubit.  The high-bias regime, where S(2,0) is the
dominant component of the ground state, is important for initialization of an ST qubit, but is not considered in this study.

\subsubsection{Beyond the Hund-Mulliken model}

Beyond the Hund-Mulliken model one can build CI calculations with various degrees of truncation.  For example, by including the single-electron
$p$ orbitals, one can introduce anisotropy into the electronic states,\cite{Hu_PRA00} which allows the electrons to spread their wave functions
better in order to minimize the total energy (therefore a better chemical bond), as compared to the HL and HM models, which are based solely on
$s$ orbitals.  The inclusion of higher orbital states also allows a better account of the correlations between the two electrons, leading again
to lower total energy.  On the other hand, it is also important to note again that a finite-basis CI calculation is not necessarily always
better than the simple HL and HM calculations.\cite{Herring_Magnetism}  Furthermore, the simple and analytical HL and HM models have the
advantage of providing transparent physical pictures.  Therefore in the present calculation we use only the HL and HM models, with particular
emphasis on identifying the range of validity of these models under various circumstances.  We note here that a CI calculation was recently done
to explore the highly biased regime where the doubly occupied $S(2,0)$ is the ground state.\cite{Nielsen_09}  This is a complementary study to
what we explore in this work, in analogy to the case of a helium atom versus a (biased) hydrogen molecule.

\subsection{Reliability of exchange calculations}
\label{sec:reliability}

The HL and HM methods used in this paper generally underestimate the electron correlations in the double dot, and results obtained using these
approaches are more appropriately thought of as order-of-magnitude estimates.  However, these methods do capture the basic physics of exchange
coupling in DQD systems and calculations performed under these approximations can serve as guidelines for initial development of silicon-based
DQD structures.  In this section we discuss the expected regime of applicability of these methods.

We first clarify the relationship between the valley splitting $\Delta$ and the other energy scales in the problem as we work within a
single-valley approximation.  To have reliable control over the initialization process, when the electrons tunnel into the quantum dot from an
outside reservoir, it is necessary for $\Delta$ to exceed the thermal broadening of the reservoir Fermi level: $\Delta \gg k_BT$.  If Coulomb
interaction can couple different valleys, it could pose additional requirements on $\Delta$.  Indeed, Coulomb interaction can connect charge
densities from different valleys, whether on the same dot or on different dots.  However, Coulomb matrix elements involving overlaps of states
from different valleys are strongly suppressed\cite{Culcer_09} due to the large wave vectors $k_{z} \! = \! \pm 0.85 \, (2\pi/a_{0})$ that
separate the valleys, where $a_{0} = 5.43$ $\AA$ is the lattice constant of silicon.  Therefore, as long as the two-electron system is prepared
in the ground valley eigenstate, neither changes in quantum dot potential (which is slowly varying in space) nor Coulomb interaction can
couple the electrons to the higher-energy valley states.  In other words, the only requirement on $\Delta$ in our calculation is $\Delta \gg
k_BT$ for initialization.

One of the main sources of problems in HL and HM calculations is the fact that the single-dot states used to build the two-electron bases can be
unreliable, especially for small interdot distances and/or for small barrier heights. This is related to the well-known failure of the HL method
when the overlap $l$ between the L and R states becomes too large. It was shown\cite{Burkard_PRB99} that when the parameter $c\! \equiv \!
\sqrt{\pi/2} \, e^{2}/(\kappa a \hbar \omega_{0}) = \sqrt{\pi/2} \, \tilde{a}$ (proportional to the ratio of the single-dot Coulomb energy to
the confinement energy) becomes larger than $2.8$ in 2D, the HL method gives the unphysical result of $J\! < \! 0$ at zero $B$ for small $d$.
This occurs also in 3D case with isotropic mass,\cite{Calderon_PRB06} only at higher values of $c\! > \! 5.8$.  Thus, above a certain size for
the quantum dots (as measured in terms of $\tilde{a}$ and determined by the ratio between Coulomb and confinement energy), the HL and HM
approaches will surely fail qualitatively below a threshold in $d$. Compared to a GaAs double dot with the same $d$ and $a$, in a Si/SiO$_{2}$
system the Coulomb energy is larger due to the reduced screening caused by the adjacent SiO$_{2}$ layer while the kinetic energy is smaller due
to the larger effective mass.  Both these changes lead to smaller $a^*_\text{B}$. Consequently, the critical value of $c$ is reached in smaller
dots in Si/SiO$_{2}$ (larger $\hbar \omega_0$ to compensate for smaller $\kappa$, from a different perspective) compared to those in GaAs,
reducing the range of dot sizes that can be accurately modeled by the HL and HM methods.

\begin{figure}
\includegraphics[width=0.99\linewidth]{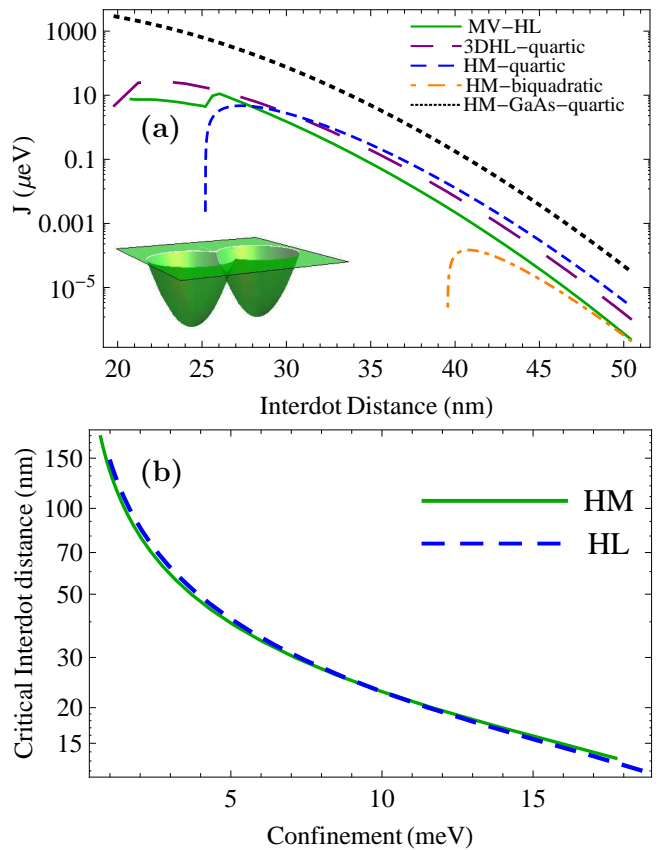}
\caption{ (a) Exchange energy as a function of interdot distance at confinement energy $8$ meV ($c \! \approx \! 4.18$, $a \! \approx \! 7.1$
nm).  The solid line corresponds to the Heitler-London calculation with the MV wave functions as in
Ref.~\onlinecite{Saraiva_PRB07} ($r  \approx 24$ nm and $\beta  \approx 0.089$). The dotted line is HM calculation in 2D GaAs system with the
same size of dot. The long-dashed line is a 3D calculation using the HL method with confinement radius $3$ nm in the $\hat{\bm z}$-direction.
The short-dashed line corresponds to the 2D HM method with the quartic potential model while the dot-dashed one is the 2D HM approach with the
biquadratic potential model. The inset shows the modified double quantum dot potential $V$ from Eq.~\ref{eq:potential}. (b) Critical interdot
distance ($2d_c$) as a function of confinement energy using the quartic potential model. The solid line corresponds to the HM method while the
dashed line corresponds to the HL method. } \label{fig:23D2D}
\end{figure}

Clearly, the case of closely-spaced dots can only be addressed by more sophisticated quantum-chemical approaches where a larger basis of
single-particle orbitals is used.\cite{Hu_PRA00,Pedersen_PRB07,Nielsen_09}  Here we choose to perform HL and HM calculations for weakly coupled
dots.  The criterion used to ascertain the qualitative applicability of these simple approaches is for the exchange splitting $J$ to increase
monotonically with decreasing $d$ at zero magnetic field.  We define the critical distance $2d_{c}$ as the interdot separation at which the sign
of the derivative $\partial J/\partial d$ changes, and $J$ starts to decrease toward zero and negative values at even smaller $d$. In
Fig.~\ref{fig:23D2D}(a) one can see that the HL and HM methods fail at values of $d_{c}$ which are in the range of possible interdot separations
in experiments.  The critical $d_{c}$ as a function of confinement energy is shown in Fig.~\ref{fig:23D2D}(b), from which we can see that for
interdot distance of $2d \! \approx \! 50$ nm we can only consider dots with $\hbar\omega_{0} \! > \! 5 $ meV, which corresponds to Fock-Darwin
radii $a \! < \! 9$ nm for Si/SiO$_{2}$.

While for $d\! > \! d_{c}$ the exchange splitting changes according to the expected trend, we still need to establish the reliability of HL and
HM calculations quantitatively.  For this purpose we have adopted an approach proposed in Ref.~\onlinecite{Saraiva_PRB07}, in which the
reliability of the HL approach was significantly improved by employing a more realistic DQD potential and choosing an appropriate variational
form of the single-electron orbitals for this potential.  The modified potential is illustrated in the inset of Fig.\ref{fig:23D2D}(a).
Analytically it is given by
\begin{eqnarray}
V=\left\{
\begin{array}{l l l l}
\frac{1}{2} m \omega^{2}_{0} \left(\text{Min} \left\{ (x-d)^2,(x+d)^2 \right\} + y^2 - r^2 \right),
\\
\\
\ \ \ \ \ \text{if} \ \ \text{Min} \left\{(x - d)^2,(x + d)^2 \right\} + y^2 < r^2
\\
\\
0, \ \ \text{if} \ \ \text{Min}\left\{(x - d)^2, (x + d)^2 \right\} + y^2 \geq r ^2
\end{array} \right.
\label{eq:potential}
\end{eqnarray}
where $r$ is the radius at which the confinement potentials on both dots are truncated, different from the Fock-Darwin radius of the dot. In our
calculations we choose $d/r$ in the range of 0.48-1.6. The depth of the dot potential that we choose in the calculations below allows at
least three single-particle energy levels to exist in each dot. Fig.~\ref{fig:23D2D}(a) shows that the potential function in
Eq.~(\ref{eq:potential}) yields exact harmonic confinement around the quantum dot centers and becomes zero at larger distances from the centers.
The single-particle wave functions have to be modified in order to reflect the fact that the potential barrier around the double dot is not
infinite.  We consider single-particle wave functions that are centered at the minimum of each dot to take the following variational form:
\begin{eqnarray}
\varphi (\rho) = \left\{
\begin{array}{l l}
A_1 \exp \left[-\frac{\beta \rho^2}{2 a^{*2}_{\text{B}}} \right] , \ \rho < \mu \nonumber
\\
\\
A_2 \exp \left[-\frac{\alpha \rho}{2 a^{*}_{\text{B}}} \right],  \ \ \rho \geq \mu
\end{array} \right.
\end{eqnarray}
where $\rho$ denotes the distance from the center of the single dot. This piece-wise wave function involves new parameters $A_1$, $A_2$, $\mu$,
$\alpha$ and $\beta$.  However, the normalization condition and the boundary conditions (continuity of the wave function and its derivative) at
$(x \mp d)^2+y^2= \mu^2$ would leave us two independent parameters, which we choose to be $\beta$ and $\mu$,  the first of them related to the
radius of the electron wave function in the dot and the second being the radius at which the change of decay behavior of the wave function
occurs. The parameters $\beta$ and $\mu$ are obtained variationally by minimizing the single-particle energy in a single dot, the potential of
which is given by
\begin{eqnarray}
V(\rho)=\left\{
\begin{array}{l l}
\frac{1}{2}m\omega^{2}_{0}( \rho^2- \mu^2), \ \rho < r \nonumber
\\
\\
0,  \ \ \rho \geq r
\end{array} \right.
\end{eqnarray}
This matched variational (MV) wave function adopted here gives a larger overlap compared to the Gaussian orbitals used previously since the
wave function has an exponential tail.  The variational procedure also allows the MV orbitals to be better matched to the potentials of the two
dots. Furthermore, the total potential at large interdot distances is now simply a sum of the two potentials corresponding to the two dots, in
analogy to the potentials of two atoms that make up a molecule, which suggests that the single-dot MV wave functions are a better starting point
for HL and HM calculations.

The same approach can be used in the case of nonzero $B$ field. For a single dot we use the gauge in which $\mathbf{A}\!=\! B(-y,x,0)/2$, which
produces an effective parabolic potential $\frac{1}{2}m\omega_{L}^2\rho^{2}$ and a term proportional to the $\hat{L}_{z}$, the projection of the
angular momentum on the $\hat{\bm z}$ axis. Since we use a trial wave function $\varphi(\rho)$ given above which depends only on $\rho$, only the
$B$-induced harmonic confinement needs to be taken into account. The calculation is analogous to the previous one.

The improvement of the results in the MV approach over the HL calculation with Gaussian orbitals obtained in Ref.~\onlinecite{Saraiva_PRB07}
is consistent with the results from Ref.~\onlinecite{Pedersen_PRB07}, where HL results for a biquadratic DQD potential were compared with a
multiorbital calculation. The MV approach was also shown to be in qualitative agreement with the analytically interpolated expressions for $J$
in the double-donor potential,\cite{Saraiva_PRB07,Ponomarev_PRB99} where one can use the exact results for $J$ obtained using the
Gorkov-Pitaevskii method.\cite{Gorkov_PRB03}

These results indicate that the MV calculation should be a good approximation to the exact result for exchange splitting, at least for larger
$d$. In Fig.~\ref{fig:23D2D}(a) we show that using the MV approach we can calculate $J$ for more closely spaced dots than in the case of HL or HM
approximation with Gaussian orbitals. The dip in $J(d)$ calculated by the MV method at $d\! \approx \! 26$ nm is most probably an artifact due
to using the HL method with the confinement potential having a kink between the dots --- a similar feature was seen in HL calculation with a
biquadratic potential in GaAs.\cite{Pedersen_PRB07} Furthermore, the values of $J$ obtained using the different methods and potentials exhibit
the same behavior at $d\! > \! d_{c}$. The spread in values of $J$ obtained with different potentials and wave functions is about an order of
magnitude, but since the $J$ itself changes by orders of magnitude with quite small relative change in $d$, this feature does not affect the
qualitative information which we can gain from our calculations. The main point here is that at $d \! > \! d_{c}$ various methods of calculating
$J$ agree, strengthening our assertion than at these interdot distances the results of HL and HM approaches are semi-quantitatively reliable. In
Sec.~\ref{sec:results} we will present the results of calculations for symmetric and asymmetric (biased) DQDs using these methods, with the
quartic potential and the modified MV potential for a symmetric DQD, and the biquadratic potential for the asymmetric DQD. We do not use the
biquadratic potential for the symmetric DQD because the unrealistic kink in the central barrier overly suppresses the interdot overlap and
biases the result in favor of the triplet state, resulting in the failure of the HM calculation at much larger value of $d_{c}$.

In Fig.~\ref{fig:23D2D}(a) we also show the result of the HL calculation with 3D Gaussian orbitals, i.e., ~the wave functions from
Eq.~(\ref{eq:phi}) multiplied by the Gaussian envelope $\exp(-z^2/2a^2_{\perp})$ in the $\hat{\bm z}$ direction.  Here $a_{\perp}$ is the
Fock-Darwin radius in the $\hat{\bm z}$ direction, which in our calculation is $a_{\perp} \! =\! 3$ nm.  As expected from the lowering of the
Coulomb energy in the 3D calculation due to the increased spread of the wave functions, the critical interdot distance $d_{c}$ is smaller in the
3D case, thus increasing somewhat the range of validity of the HL and HM approaches.\cite{Calderon_PRB06} However, since the confinement in the
$\hat{\bm z}$-direction is generally stronger than in the $xy$-plane and the Fock-Darwin radius associated with it smaller, making the spatial integrals
anisotropic, we cannot obtain a closed-form expression for the Coulomb contribution to the exchange interaction.

For comparison we have also performed the calculations for GaAs using the same interdot distance and the same Fock-Darwin radii $a$ as the
corresponding Si dots (an example is shown in Fig.~\ref{fig:23D2D}). The resulting exchange splittings are typically an order of magnitude
larger than the ones obtained in Si.  However, since decreasing $d$ by about $10\%$ allows one to gain an order of magnitude in $J$, there is no
practical obstacle in obtaining the same values of $J$ in Si and GaAs.

In summary, in this section we have shown that the HL and HM methods are valid in smaller parameter regimes in a Si/SiO$_2$ system as
compared to GaAs DQDs due to the enhanced Coulomb interaction and larger effective mass in Si/SiO$_2$.  These are facts one needs to be aware of
when calculating electronic states in Si.  One consequence for experiment is that the QDs in Si have to be made smaller and closer in order to
have large exchange splittings. From comparison of calculations with different methods and different model potentials we draw the conclusion
that the HL and HM calculations are semiquantitatively accurate at moderately large interdot distances.

\section{Spin qubit architectures in quantum dots}
\label{sec:architectures}

Our study of electron exchange in a DQD is mainly motivated by two approaches to spin quantum computation using quantum dots.  In the original
spin QC proposal, which we refer to as Loss-DiVincenzo (LDV) architecture,\cite{Loss_PRA98} an array of quantum dots is used as a set of spin
qubits.  Each dot is singly occupied and the spin of the electron in each dot constitutes the qubit.  Single-qubit operations can be implemented
by rotating the single spins using a magnetic field.  Two-qubit operations, which are needed in order to implement logical gates, can be
realized by means of a pulsed (gate-controlled) exchange interaction between neighboring dots by lowering the inter-dot potential barrier and
allowing the electron wave functions to overlap.  Remote two-qubit operations can be realized by swapping the two qubits next to each other,
again assisted by pulsing on and off exchange interaction between neighboring quantum dots.

The LDV scheme involves a system of symmetric dots since, in general, no electrical bias is applied between the adjacent dots at any time.  Indeed
small bias about the symmetric point will only reduce the exchange splitting $J$.\cite{Burkard_PRB99}  The model potential is symmetric and the
best choice for the two-dot scenario is the quartic potential introduced in Sec. \ref{sec:formalism}.

The simplicity and elegance of the LDV architecture have motivated extensive experimental studies of single spin properties such as coherent
control, coherence, and measurement.\cite{Ono_Science02, Elzerman_Nature04, Johnson_Nature05, Koppens_Science05, Hanson_RMP07,
MarcusGroup_NatureNano07, Nowack_Science07, Shaji_NP08, Koppens_PRL08}  The experimental problems encountered in these explorations, such as
difficulties in reliable and fast spin initialization and single spin ESR control, have led to theoretical proposals where pure electrical
control of encoded spin states can be used for universal quantum computation.\cite{DiVincenzo_Nature00, Levy_PRL02}  In particular, a series of
experiments\cite{Johnson_Nature05, Petta_Science05, Reilly_Science08, Pioro_NatPhys08} have shown the viability of two-spin encoding using spin
singlet and unpolarized triplet states.

For a singlet-triplet qubit, the exchange splitting $J$ plays the same role as the Zeeman splitting does for a single-spin qubit.  Turning it on
produces a phase difference between the singlet and triplet states so that different superposition states can be realized.  For the
experimentally realized ST qubit,\cite{Petta_Science05,Taylor_PRB07} the double dot is in a strongly biased regime so that (1,1) and (2,0)
singlet states are close to resonance in the absence of tunneling. The exchange splitting is turned on by starting deep in the (1,1) regime
[where (1,1) singlet and triplet states are nearly degenerate ground states of the DQD, indicating that the interdot tunnel coupling is small]
and increasing the mixing between the (1,1) and (2,0) singlet states due to tunneling.\cite{Petta_Science05,Taylor_PRB07}  In other words, the
exchange splitting here is dominated not by the Coulomb interaction but by tunneling-induced mixing between the levels detuned by $\epsilon$.

One of the potential drawbacks of working with a highly biased double dot is that charge noise affects the bias linearly, which in turn leads to
fluctuations in the exchange splitting $J$.\cite{Culcer_APL09}  This amounts to a dephasing mechanism for the singlet-triplet qubit if $J$ is
finite. Therefore in the current study we also investigate whether it is possible to reduce the sensitivity of this system to charge
fluctuations.

\section{ Results}\label{sec:results}

Based on the above discussions of the reliability of the HL and HM approaches within the specific regime we are considering,  we calculate the
exchange splitting between the triplet and singlet states in a coupled DQD near a Si/SiO$_2$ interface.  Our calculations are done for
experimentally achievable sizes of dots.  We investigate the dependence of exchange on the confinement energy, interdot distance, and the
magnetic field in the symmetric (Loss-DiVincenzo) case, and the sensitivity of the exchange coupling of the two LDV qubit to charge noise.  For
the asymmetric case (singlet-triplet qubit), we determine the dependence of the exchange coupling on the interdot detuning within the
Hund-Mulliken model, and we discuss the presence of the optimal points with respect to different kinds of charge noise.

\subsection{Loss-DiVincenzo architecture}

Our discussion of the Loss-DiVincenzo architecture relies on the quartic model potential for reasons discussed in Sec.~\ref{sec:reliability},
where we showed that this potential gives physical results for $J$ in a wider range of interdot distances compared to the biquadratic model.

\begin{figure}
\includegraphics[width=0.99\linewidth]{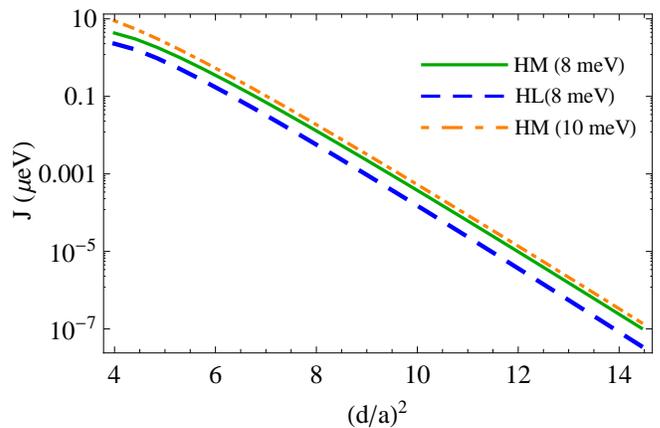}
\caption{(color online). Exchange coupling as a function of $\frac{d^2}{a^2}$ using the quartic model potential.  The solid and dashed lines
correspond to HM and HL with $\hbar \omega_0 \!=\!8$ meV ($a \! \approx \! 7.1$ nm) respectively while the dot-dashed line corresponds to HM
with $\hbar\omega_0 \! = \!$ 10 meV ($a \! \approx \! 6.3$ nm). } \label{fig:exchange_versus_d}
\end{figure}

In Fig.~\ref{fig:exchange_versus_d} we plot $J$ as a function of $\frac{d^2}{a^2}$ at zero magnetic field.  The results here are straightforward
to understand.  As the interdot separation increases, the overlap $l\!=\! \exp (-d^2/a^2)$ decreases as a Gaussian.  Correspondingly the
exchange coupling $J$ decreases according to Eq.~(\ref{eq:J_HL}), and the decrease is dominated by the $l^2$ prefactor for $J$.
Fig.~\ref{fig:exchange_versus_d} shows that with a confinement energy of $8$-$10$ meV, exchange can reach $1 \mu $eV in the range of interdot
distance of about 30 nm.  In experiment, the confinement energy will typically be smaller than $8$ meV, which implies that the actual exchange
coupling at the Si/SiO$_2$ interface could be orders of magnitude larger. We note that $J \! \sim \! 1$ $\mu$eV is already well within energy
resolution of current generation experiments,\cite{Petta_Science05}.  It corresponds to a gating time of about a nanosecond.  The coherence time
of a single spin in GaAs quantum dot has been measured in a spin echo experiment to be about $300$ ns at low magnetic field ($B \! < \! 0.1$
T),\cite{Koppens_PRL08} and longer coherence times are expected at higher $B$ fields.\cite{Witzel_PRB06, Yao_PRB06, Deng_PRB06, Deng_PRB08,
Coish_PRB08, Cywinski_PRL09, Cywinski_PRB09} Consequently, even a $J$ of 1 $\mu$eV magnitude should be enough for quantum-computation purposes
within the LDV and ST architectures.

\begin{figure}
\includegraphics[width=0.99\linewidth]{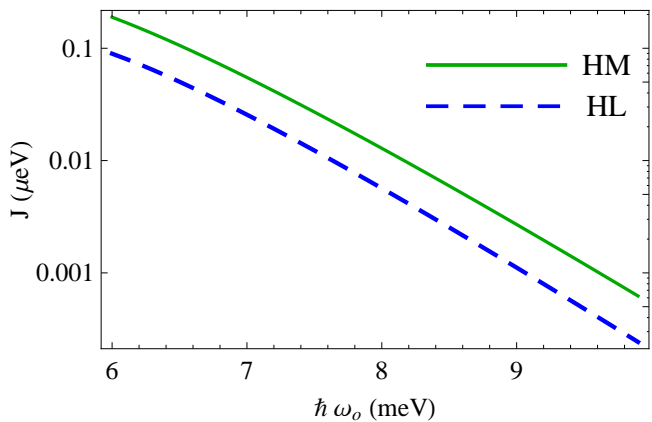}
\caption{(color online). Exchange coupling as a function of confinement energy using the quartic model potential for two dots separated by $40$
nm.  The solid line corresponds to the HM approach while the dot-dashed line corresponds to the HL method. } \label{fig:sy_exchange_VS_cf}
\end{figure}

Figure~\ref{fig:sy_exchange_VS_cf} shows $J$ as a function of confinement energy $\hbar\omega_{0}$. Increasing the confinement energy changes
the central barrier height ($V_{b} \! = \! \frac{1}{8}m\omega^{2}_{0}d^2$) between the two dots,  but more importantly it squeezes the
Fock-Darwin radius since $a = \sqrt{\hbar/m\omega_0}$, which has a similar effect on the exchange coupling as varying the interdot distance.
When the confinement energy is larger than 10 meV (with $a < 6.5$ nm), the exchange coupling becomes extremely small at interdot distances above
$30$ nm since $l^2$ now is smaller than $10^{-10}$.  On the other hand, when the confinement energy is smaller than $6$ meV (with $a > 8$ nm),
we obtain an unphysical triplet ground state for interdot distances smaller than $32$ nm.  The reason for this unphysical behavior, which has also
been observed for interatomic calculations before, lies in the failure of HL and HM models to properly account for electron
correlations.\cite{Herring_Magnetism}

\begin{figure}
\includegraphics[width=0.99\linewidth]{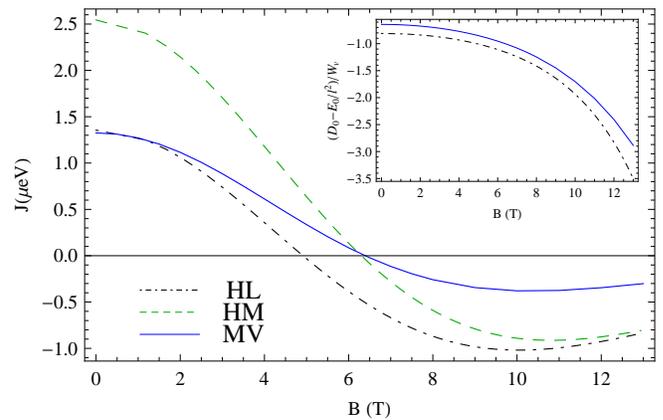}
\caption{(color online).  Exchange coupling as a function of magnetic field  at confinement energy $8$ meV and interdot distance $30.24$ nm. The
solid line corresponds to the MV-HL calculation, and the dashed (dot-dashed) lines are HM (HL) calculations with the quartic potential. The
inset presents the value of $(D_{0}-E_{0}/l^2)W_{v}$ [see Eq.~(\ref{eq:J_HL})] as a function of magnetic field, which indicates the relative
importance of Coulomb interaction compared to single-particle kinetic and potential energy terms.}
\label{fig:sy_magnetic1}
\end{figure}

In Fig.~\ref{fig:sy_magnetic1} we plot $J$ as a function of a uniform applied magnetic field $B$ in the $\hat{\bm z}$ direction. The DQD system
shows a clear transition from an antiferromagnetic (singlet) ground state at low field to a ferromagnetic (triplet) ground state at high field.
This trend, which appears here in (single-valley) Si DQDs, is consistent with previous results obtained for GaAs DQDs.\cite{Burkard_PRB99,
Hu_PRA00, Pedersen_PRB07, Scarola_PRA05}  In essence, as the external field increases, the interdot overlap decreases so that the overall
magnitude of $J$ goes down.  However, the quantum interference brought about by the magnetic phases in Eq.~(\ref{eq:phi}) causes the magnitude
of the exchange Coulomb term to increase relative to the direct Coulomb term (as shown in the inset of Fig.~\ref{fig:sy_magnetic1} for the HL
and MV-HL calculations, where $E_0/l^2 - D_0$ increases monotonically with the applied magnetic field.  The singlet-triplet crossing occurs at
the field where the quantity plotted in the inset reaches the value of -1), so that at larger fields the triplet configuration is favored
energetically, as shown in Fig.~\ref{fig:sy_magnetic1}. More involved numerical calculations performed for GaAs DQD have shown that further
changes in the two-electron ground state are possible at higher fields.\cite{Scarola_PRA05} However, whether the triplet will remain the ground
state at even higher magnetic fields cannot be determined by our simple model calculation---indeed the inset of Fig.~\ref{fig:sy_magnetic1}
seems to indicate that there will be no further singlet-triplet crossings at higher fields. The exchange coupling will go to zero asymptotically
with increasing $B$, since the interdot overlap becomes so small at large $B$ that the two dots are effectively fully separated.

\begin{figure}
\includegraphics[width=0.99\linewidth]{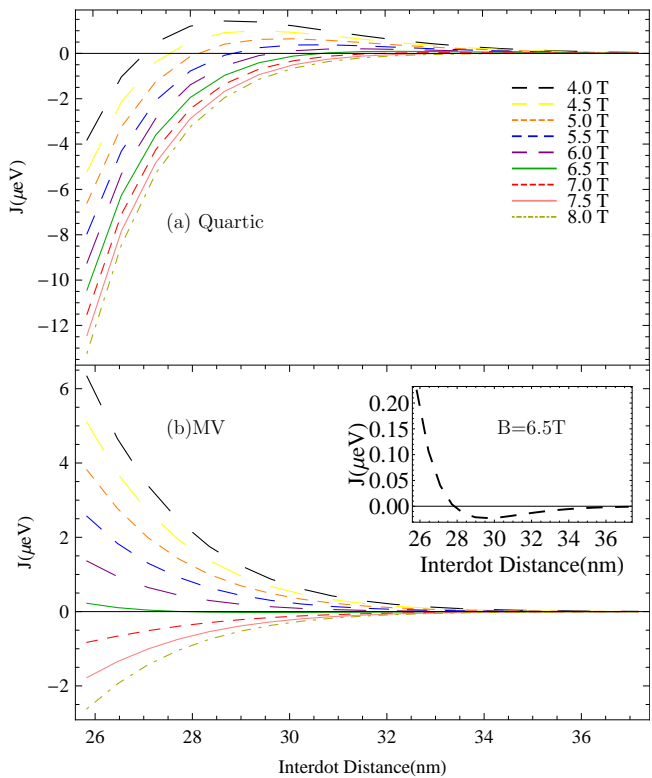}
\caption{(color online). Exchange coupling as a function of interdot distance at confinement energy $8$ meV with different magnetic field. Fig. (a) shows the result of quartic model. Fig. (b) shows the result of MV model($r  \approx 24$ nm) The inset shows the result of exchange coupling as a function of interdot distance at $B=6.5$ T.} \label{fig:sy_magnetic_VS_d1}
\end{figure}

Figure~\ref{fig:sy_magnetic_VS_d1} shows $J$ as a function of the interdot distance $d$ at finite $B$. The top panel shows the results obtained
using the quartic model potential and HM method, while the bottom panel shows the results obtained using the MV approach. As shown in
Fig.~\ref{fig:sy_magnetic1}, for a given $d$ the exchange coupling $J$ changes its sign with increasing $B$ at specific interdot distances.
Changing $d$ while keeping $B$ fixed again changes the magnetic phase factor, leading to possibly nonmonotonic dependence of $J$ on $d$.
Interestingly, the behavior of $J$ as a function of $d$ is different in these two models. When the MV approach is used, the magnitude of $J$ is
monotonically decreasing with increasing dot separation, with the exception of a small range of $B$ around 6.5T (shown in the inset). Within
this range the curve of $J$ versus $d$ is at first positive, then drops below zero and goes through a minimum. For these few values of $B$ there
is thus a point on the curve where $\partial J/\partial d = 0$. On the other hand, when the quartic model potential is used, $J$ is negative at
smaller $d$ and relatively low $B$, and it has a maximum over a wide range of $B$. Since the negative $J$ at smaller $d$ in the quartic
potential calculation can actually be due to the failure of the HM method at small interdot distances, and in view of the wider range of
applicability of the MV method (see Fig.~\ref{fig:23D2D}), we expect that the trend shown by the $J(d)$ curve produced by the MV method as the
magnetic field increases is more likely to match realistic experimental situations, while caution must be exercised in interpreting results
obtained using the quartic model potential.

When two electron spins are exchange coupled, the spins are exposed to the electrostatic environment. Fluctuations in  the electrostatic
potential will lead to fluctuations in the exchange splitting $J$, which in turn leads to gate errors.\cite{Hu_PRL06}  In general we can write
$\Delta J = \sum_i (\partial J/\partial V_{i})\, \Delta V_{i}$, where the partial differentiations reflect the sensitivity of the double dot
system to the particular fluctuation, and the $V_{i}$ are different sources that contribute to the barrier or the interdot
bias.\cite{Culcer_APL09} For example, $\partial J/\partial V_b$, with $V_b$ being the barrier in the model potential, should tell us how much
voltage fluctuations in gate electrodes that define the DQD could affect the exchange, and it is important to study $\partial J/\partial V_b$
and ways to minimize it.

\begin{figure}
\includegraphics[width=0.99\linewidth]{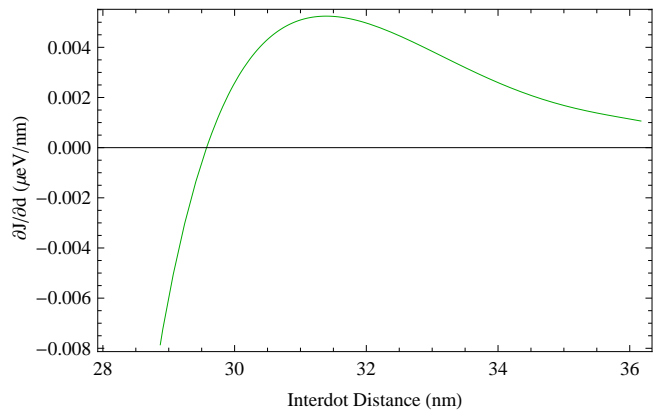}
\caption{(color online). $\partial J/\partial d$ (the derivative of $J$ with respect to the interdot distance) as a function of interdot
distance at a confinement energy of $8$ meV and at $B\! = \! 6.5$ T, using the MV approach within the HL method. }
\label{fig:sy_magnetic_VS_d2}
\end{figure}

While the interdot distance in the model quartic potential is directly related to the central barrier height $V_b \! =\! m \omega_0^2d^2/8$,
such a simple relation generally cannot be established for more realistic potentials.  It certainly does not hold for the truncated potential
used in the MV calculation, the results of which should be more reliable.  We would like to point out that the derivative $\partial J / \partial
d$ quantifies the response of $J$ to fluctuations of the external potential which predominantly cause the change in the interdot distance,
i.e., ~they are even with respect to $yz$ plane.  The inset of Fig.~\ref{fig:sy_magnetic1} indicates that there is an optimal point (at which
$\partial J / \partial d\! =\! 0$) with respect to such fluctuations predicted by the MV calculation at $B\!= \! 6.5$ T. The corresponding plot
of $\partial J / \partial d$ is shown in Fig.~\ref{fig:sy_magnetic_VS_d2}. Outside the small range of magnetic field around 6.5 T, the magnitude
$|\partial J/ \partial d|$ is simply monotonically decreasing with increasing $d$.  We note that $\partial J/\partial d$ in the absence of a
magnetic field shows a qualitatively similar behavior as those with $B \neq 6.5$ T and there is no optimal point in that case.

\subsection{Singlet-triplet qubit}
The description of the ST qubit necessarily requires taking into account the doubly occupied state in one dot, since tunneling between (1,1) and
(2,0) singlet states is the physical process that lowers the energy of the singlet relative to the triplet state, as we discussed in
Sec.~\ref{sec:architectures}.  Therefore this problem can only be tackled using the HM approximation.  In modeling a biased DQD we make use of
the biquadratic potential with an electric field added along the $\hat{\bm x}$ direction.  This choice is due to the fact that the curvature of
the quartic potential is altered considerably by the addition of the electric field, and the minima at the centers of the two dots have rather
different shapes, as can be seen in Fig.~\ref{fig:asymmetric}.  On the other hand, the problems with the biquadratic potential in the symmetric
case (i.e.~the failure to obtain $J\! > \! 0$ occuring at interdot distances larger than when using the quartic potential) are not as severe in
the biased case: the physics which controls $J$ in the regime of parameters relevant for the operation of the ST qubit is not qualitively
sensitive to the details of the potential shape.

\begin{figure}
\includegraphics[width=0.99\linewidth]{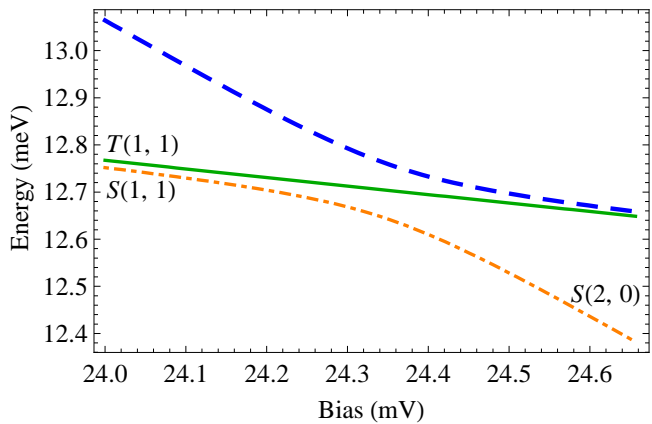}
\caption{(color online). Energy spectrum as a function of bias near the anticrossing. The confinement energy is $6$ meV and E=0.6 V/$\mu$m.  We
have used the asymmetric biquadratic potential. } \label{fig:energy}
\end{figure}

\begin{figure}
\includegraphics[width=0.99\linewidth]{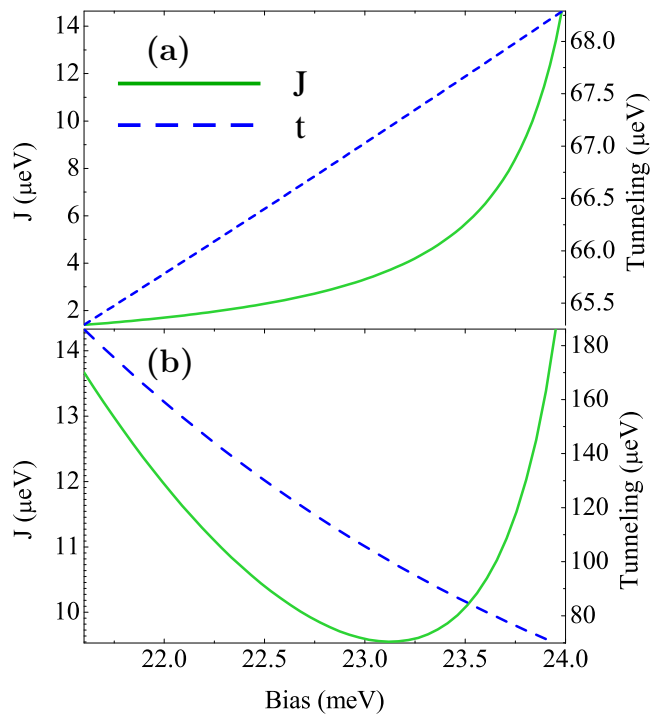}
\caption{(color online). Exchange coupling (solid line) and tunnel coupling (dashed line) versus the bias near the (1,1)-(2,0)-singlet resonant
point.  The upper panel (a) presents the results when change of bias is by varying the electric field, with the interdot distance fixed at 40 nm
and confinement energy $\hbar\omega_0$ = 6 meV.  The lower panel (b) presents the results when change of bias is by varying the interdot
distance $d$ with the electric field fixed at E=0.6 V/$\mu$m and confinement energy $\hbar\omega_0$ = 6 meV.} \label{fig:Graph1}
\end{figure}

The exchange coupling in a biased DQD is the energy difference between the two lowest eigenvalues of the HM Hamiltonian in Eq.~(\ref{eq:H_HM}).
In Fig.~\ref{fig:energy} we show the three lowest eigenvalues of the HM Hamiltonian as a function of the detuning energy near the
$(1,1)$-singlet to $(2,0)$-singlet anticrossing. The evolution of the DQD spectrum near this charge transition is parametrized by the detuning
$\epsilon \! = \! 2 (d/a) f \hbar \omega_0 \! = \! 2eEd$ [see Sec. II.A and Eq.~(10)], which is a linear function of both the interdot
distance and the electric field. For relevant experimental manipulations it is natural to assume that the change in electric field is the main
driving force behind changes in detuning. However, one could also keep $f \! \sim \! E$ fixed, and vary $d$ in order to vary the detuning. As
$\epsilon$ increases in Fig.~\ref{fig:energy}, the ground state of the DQD changes gradually from the $(1,1)$ spin-singlet state to the $(2,0)$
spin-singlet state. The anticrossing point at which $\epsilon \! = \! \epsilon_{c}$ can be viewed, approximately, as the point where the
composition of the ground state changes from predominantly $(1,1)$ to mostly $(2,0)$.  We focus on the spectrum near this anticrossing,
specifically for $\epsilon \! < \! \epsilon_{c}$, which is the range of detuning used in qubit manipulations [the regime where $S(2,0)$ is the
ground state is used only for initialization of the ST qubit]. For these values of $\epsilon$ we calculate the exchange coupling, tunnel
coupling, and we investigate the sensitivity to charge noise of the asymmetric DQD.  In the following we use a confinement energy of 6 meV,
where $c \!= \! 4.84$, the Fock-Darwin radius $a \!= \!8.16$ nm and $d/a >2$. The exchange coupling in the biased DQD can be then tuned in a wide
range ($0 \sim 100 \mu eV$) near the anticrossing point.

Fig.~\ref{fig:Graph1} shows the exchange splitting $J$ and tunnel coupling $t$ as functions of $\epsilon$ near the anticrossing point. The
exchange splitting depends on both tunnel coupling and detuning, and it becomes very large near the anticrossing (resonant) point between the
$S(1,1)$ singlet and the $S(2,0)$ states.  In the upper (lower) panel we show the results when detuning is changed via variation of applied
electric field (interdot distance).  Comparison of the two panels shows that the electric field and the interdot distance have quite different
effects on $J$ and $t$.  As the electric field increases (upper panel), the tunnel coupling increases slightly (less than 5\% in the given range
of detuning) because the barrier height relative to one of the dots decreases slightly.  In addition, the overlap between neighboring orbitals
does not depend on the electric field, so that the Coulomb terms do not change at all, and $t$ is only affected by slightly different barrier
height.  Therefore, the change in $J$ in this case (varying by a factor of 7) is dominated by the change in the detuning, since $J\! \sim \! 2t^2/(\epsilon_{c}-\epsilon) $ in the regime where S(1,1) is the dominant component of the ground state, away from the anticrossing point.
Conversely, while increasing the interdot distance leads to a larger $\epsilon$ as well (lower panel), it also lowers the overlap between the
left and right orbitals and exponentially reduces the tunnel coupling.  The exchange coupling in this case is determined by the competition
between the energy level detuning and the tunnel coupling.  The curve of  $J$ as a function of bias is no longer monotonous but decreases first
and then rises as shown in the lower panel of Fig.~\ref{fig:Graph1}. The presence of this extremum in $J(d)$ dependence leads to the existence of an optimal point which we discuss below.

\begin{figure}
\includegraphics[width=0.99\linewidth]{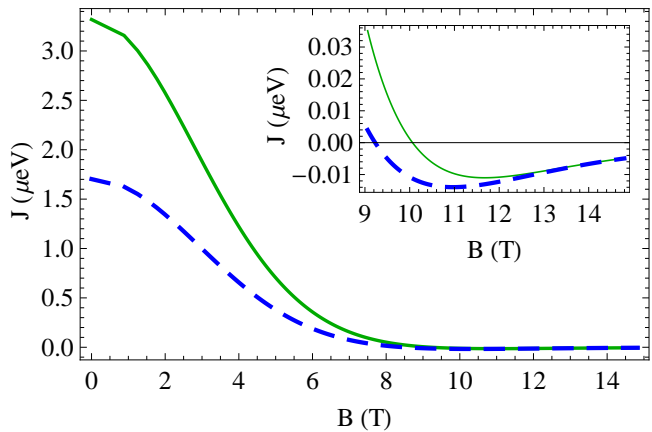}
\caption{(color online) Exchange coupling as a function of magnetic field for two values of the bias detuning $\epsilon$ with confinement energy
of $6$ meV and interdot distance of $40$nm.  The dashed line corresponds to $\epsilon = 22$ meV while the solid line corresponds to $\epsilon =
23$ meV.  The inset gives an enlarged version of $J (B)$ at large magnetic field.} \label{fig:m}
\end{figure}

Figure~\ref{fig:m} shows $J$ as a function of $B$ at different detunings. The overall decrease in $J$ is again due to the magnetic squeezing of
the orbital states and suppression of interdot overlap, as in the symmetric dot case.  There is also a transition from antiferromagnetic
(singlet) to ferromagnetic (triplet) spin coupling for the ground state, yet this feature is far less dramatic than in the symmetric case.  The
expanded view in the inset shows that the exchange splitting switches sign from positive to negative when the magnetic field exceeds $9$ T although the negative $J$ is very small in magnitude.  A possible reason for this suppressed singlet-triplet crossing is that in a biased DQD
the exchange is dominated by the variations in detuning and tunnel coupling.  While Coulomb interaction modifies these factors, it is not as
important as in a symmetric DQD.  Therefore, the effect of competition between direct and exchange Coulomb integrals is also less prominent.

\begin{figure}
\includegraphics[width=0.99\linewidth]{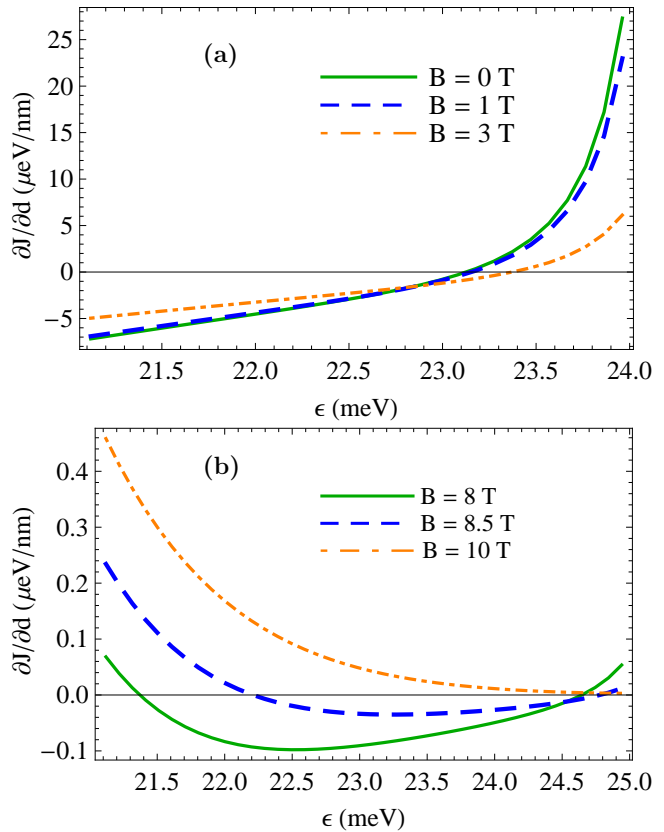}
\caption{(color online) $\partial J/\partial d$ as a function of the bias $\epsilon$ by varying interdot distance, with a fixed electric field
of $E = 0.6$ $V/\mu$m and a single-dot confinement energy of $\hbar\omega_0 = 6meV$.  Panel (a) shows one sweet point $\partial J/\partial
d \equiv 0$ at small magnetic field($B < 6$ T). While Panel (b) shows that there are two sweet spots around $B = 8$ T but no sweet point
for $B > 10$ T.}
\label{fig:d2}
\end{figure}

In an asymmetric DQD the effect of background charge noise on $J$ is governed by the $\partial J/\partial d$ and $\partial J/\partial E$
derivatives.  The first of them parameterizes the influence of fluctuations in the DQD potential that are \emph{even} with respect to $yz$ plane,
while the second controls the influence of the \emph{odd} changes. In Fig. \ref{fig:d2} $\partial J/\partial d$ is plotted at relatively low and
relatively high magnetic fields as a function of bias $\epsilon$ (the interdot distance $2d$ is varied). Panel (a) shows $\partial J/\partial d$
versus the bias detuning at small magnetic fields.  There is a sweet spot on this graph at which $\partial J/\partial d = 0$, which is also seen
in the absence of a magnetic field.\cite{Stopa_NL08}   This is a point where exchange splitting has a local minimum and the reason is the
competition between a decreasing tunnel coupling and a detuning that is approaching (1,1)-(2,0) resonant point which we discussed before (see
also the lower panel of Fig.~\ref{fig:Graph1}).  On the other hand, in panel (b) we show the results at higher $B$ fields.  At $B \! \approx \!
8$ T a second optimal point appears.  The exchange splitting has its local maximum at this point, which is due to the contribution of magnetic
phase. If the magnetic field is larger than 10 T, $\partial J/\partial d$ is always positive and has its minimum value near the anticrossing.

\begin{figure}
\includegraphics[width=0.99\linewidth]{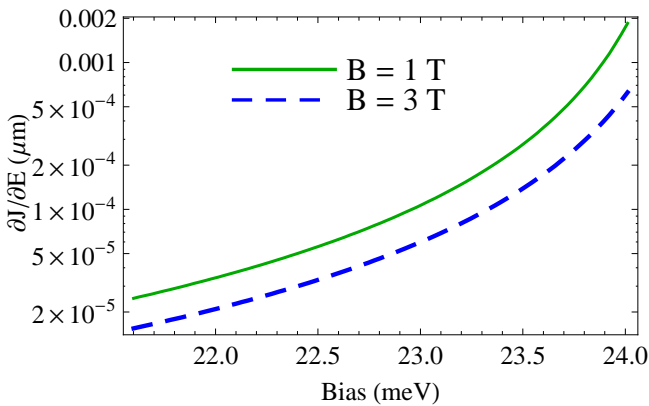}
\caption{(color online). $\partial J/\partial E$ versus the bias for different magnetic fields. The bias is varied by altering the electric
field E.  The solid line corresponds to $B=1$ T while the dashed one corresponds to $B=3$ T. }
\label{fig:d3}
\end{figure}

In Fig.~\ref{fig:d3} we show $\partial J/\partial E$ as a function of detuning at an interdot distance of $40$ nm.  The sensitivity to charge
noise, encapsulated by $\partial J/\partial E$, increases monotonically as the bias increases, and there is no optimal point.  The physical
reason for this monotonic dependence is straightforward: as bias increases, the charge-distribution difference between the singlet and triplet
states acquires a stronger dipolar characteristic [in the (1,1) configuration this difference is electric quadrupolar], so that the DQD system
becomes more susceptible to environmental charge fluctuations.\cite{Ramon_09}

We note here that while clear separation of physical effect that changing $E$ and $d$ have on $J$ is a feature specific to the model biquadratic
potential, in which the applied electric field does not change the interdot distance, our result highlights the fact that the dependence of $J$
on the potential fluctuations can exhibit local extrema if the relevant potential fluctuation influences the effective interdot distance.

\section{Summary and Conclusions}
\label{sec:summary}

We have calculated exchange coupling of two electrons in double quantum dot structures in a Si/SiO$_2$ heterostructure within the single-valley
approximation. This approximation is valid as long as the valley splitting $\Delta \gg k_BT$, which is necessary for reliable initialization of
electron-spin states for the purpose of quantum computation. Having analyzed the reliability of Heitler-London and Hund-Mulliken models for
calculating exchange splitting in a Si double dot, we investigated symmetric and asymmetric double quantum dots of experimentally achievable
sizes, with our focus on two-electron exchange splitting and its sensitivity to charge noise. Throughout this work we have also used a matched
variational approach\cite{Saraiva_PRB07} to extend the range of applicability of our results.

For the Loss-DiVincenzo architecture we find that, for the same interdot separation as in GaAs, exchange is generally smaller because of the
larger effective mass and the resulting smaller wave-function overlap.
In the presence of a magnetic field, we find that for different confinement potentials the trend exhibited by $J$ as a function of $d$ as the
magnetic field increases, and thus the existence and location of sweet spots, is different. Modeling symmetric double quantum dots in a
magnetic field with HL and HM methods requires a thoughtful choice of model potential so as to avoid unphysical results. We find that optimal
point (at which the sensitivity to charge noise is reduced) exists for $J$ as a function of $d$ for a small range of magnetic fields.

For singlet-triplet qubits, we find that the behavior of the exchange as a function of electrical bias and of magnetic field is qualitatively
similar to GaAs. We identify an optimal point at $B\!= \! 0$. We find that the sensitivity to the noise depends on whether the potential
fluctuations are changing the interdot distance $d$ or not, i.e.~whether they are odd or even with respect to the $yz$-planes.

Future work ought to include a Poisson-Schr\"{o}dinger modeling of the actual Si DQD structures, yet this must await the successful fabrication
of Si DQD structures so that more details of the gate-induced electrostatic confinement associated with the lithographic processing become known
quantitatively.  It will also be useful to go beyond our simple Heitler-London and Hund-Mulliken theories of the exchange coupling using quantum-chemical configuration interaction corrections (Ref.~\onlinecite{Nielsen_09} is an example along that line) although we do not expect any
qualitative modifications of our results with more sophisticated numerical computations.  One might also consider the possibility, as was done
for GaAs DQD structures,\cite{Hu_PRA01} that the individual dots in the Si DQD structure may have more than one electron each to determine
whether the limit of one-electron per dot is essential in Si qubit architectures or any small odd number of electrons per dot could work under
experimental conditions. Finally, the question of what happens if the valley splitting is not large (i.e. compared to the temperature) is an
important issue as discussed recently in Ref.~\onlinecite{Culcer_09}, and must also be dealt with in a fully quantitative manner.

\begin{acknowledgments}
We thank B. Koiller, M.P.~Lilly, A.L.~Saraiva, and S.N.~Coppersmith for stimulating discussions.  This work is funded by the LPS-NSA.  {\L}C was
partially supported by the European Union within European Regional Development Fund, through grant Innovative Economy (POIG.01.03.01-00-159/08,
``InTechFun'').

\end{acknowledgments}

\appendix
\section{Matrix elements in the Hund-Mulliken approach}
\label{app:elements}

In the following, we list the matrix elements in Eq.~(\ref{eq:H_HM}) and Eq.~(\ref{eq:matrix}) in units in which the energy is measured in
$\hbar\omega_{0}$ and length is measured in Fock-Darwin radius $a \! = \! \sqrt{\hbar/m \omega_0}$. The single-dot harmonic well-ground
eigenstates $\varphi _{L/R}$ for dots centered at $\pm d$ are given by Eq.~(\ref{eq:phi}). The dimensionless parameters are then: $b \! = \!
\sqrt{1+\omega_L^2/\omega_0^2}$ (related to relative strength of magnetic confinement versus harmonic confinement), where $\omega_L$ is the Larmor
frequency $\omega_L \!= \! eB/2mc$; $c\! \equiv \! \sqrt{\pi/2}e^{2}/(\kappa a \hbar\omega_{0})$ (the ratio of Coulomb to confinement energy);
and $d_0=d/a$ (the ratio of half interdot distance to Fock-Darwin radius).  In addition, the overlap is given by $l\! \equiv\! \langle L | R
\rangle = \exp\left[ d_0^2(1/b - 2b)\right]$; and the parameter associated with the applied electric field is $f \! = \! eEa/\hbar\omega_{0}$.

In order to convert the expressions below into the units of Ry$^{*}$ and a$^{*}_{B}$ one has to multiply all the matrix elements by
$\hbar\omega_{0} = 2/\tilde{a}^2$ (note also that $c = \sqrt{\pi/2} \ \tilde{a}$).

The matrix elements in Eq.~(\ref{eq:H_HM}) are as follows:

\begin{eqnarray}
\begin{array}{l l l l l l l l l l}
\varepsilon_{L/R}=\dfrac{d_0}{2\left(1-l^2\right)} \Bigg\{ \pm
\dfrac{1}{{\sqrt{b\pi }}}(e^{-b (d_0+f)^2} (-1+e^{4 b d_0 f}) )
\\
\\
\mp (d_0+f) \ $\text{erf}$\left[\sqrt{b} (d_0+f)\right] \pm (d_0-f)
\ $\text{erf}$ \left[\sqrt{b} (d_0-f)\right]\Bigg\}
\\
\\
+\dfrac{1}{\left(1-l^2\right)}\Bigg\{b+d_0^2-\dfrac{f^2}{2}-\dfrac{d_0
e^{-b (d_0+f)^2}}{\sqrt{b \pi }}
\\
\\
+\frac{1}{2} l^2 \big[-2 b+f^2+\dfrac{2 d_0 e^{-b f^2}}{\sqrt{b \pi }}+2
d_0\  f \ $\text{erf}$ \left(\sqrt{b} f\right)\big]
\\
\\
-d_0(d_0 \mp f)\ $\text{erf}$ \left[\sqrt{b}(d_0+f)\right]\Bigg\}
\\
\\
+\dfrac{d_0}{2\sqrt{\left(1-l^2\right)}} \big\{\pm \dfrac{1}{\sqrt{b
\pi }}[e^{-b (d_0+f)^2} \left(-1+e^{4 b d_0 f}\right)]
\\
\\
\pm (d_0-f) \ $\text{erf}$ \left[\sqrt{b} (d_0-f)\right] \mp (d_0+f)
\ $\text{erf}$ \left[\sqrt{b} (d_0+f)\right] \big\},
\end{array}
\end{eqnarray}

where $\text{\text{erf}} (z)$ is the error function given by $\text{\text{erf}}(z)=\frac{2}{\sqrt{\pi }}\int _0^z e^{-t^2}dt$. The
detuning is given by $\epsilon \! = \! \epsilon_R - \epsilon_L$:
%
\begin{eqnarray}
\begin{array}{l c c}
\varepsilon =\dfrac{d_0}{\sqrt{b \pi (1-l^2)}}\{e^{-b
(d_0+f)^2}-e^{-b (d_0-f)^2}
\\
\\
\ \ \ \ \ \ +\sqrt{b \pi } \{(-d_0+f) \ $\text{erf}$[\sqrt{b}
(d_0-f)]
\\
\\
\ \ \ \ \ \ +(d_0+f) \ $\text{erf}$[\sqrt{b}
(d_0+f)]\}\},
\end{array}
\end{eqnarray}
%

All the matrix elements of Coulomb interaction in the orthogonalized
(Wannier) basis can be expressed using four integrals involving the
original $\varphi_{\pm}$ states (we use the $\pm$ notation
interchangeably with $R/L$). The confinement potential-related
contribution $W_{v}$ defined in Eq.~\ref{eq:Wv} in quartic and
biquadratic models are given by the following formulas respectively;
\begin{eqnarray}
W_{v} & = &  \frac{3}{4b} \left(1+b d_0^2\right) \,\, ,
\\
W_{v}& = & \frac{2 d_0}{{\sqrt{b \pi}}} (1-e^{-b d_0^2})+ 2 d_0^2
\big(1- \ \text{erf} [\sqrt{b} d_0 ] \big)
\end{eqnarray}

Apart from the direct and exchange integrals $D_{0}$ and $E_{0}$
from Eqs~(\ref{eq:D0_def})-(\ref{eq:E0}):
\begin{eqnarray}
D_{0} & = & c\sqrt{b} \, e^{-b d_0^2} \, I_{0}( b d_0^2) \,\, , \\
E_{0} & = & l^2 \, c\sqrt{b} e^{d_0^{2}(b-\frac{1}{b})} I_{0}(
d_0^{2}(b-\frac{1}{b}) ) \,\, ,
\end{eqnarray}
we have the on-site Coulomb repulsion term:
\begin{eqnarray}
U_0 & = & \langle L(1)L(2) | \hat{C} |L(1)L(2)\rangle \nonumber \\
& = &  \langle R(1)R(2) | \hat{C} |R(1)R(2)\rangle \nonumber \\
& = & c\sqrt{b} \,\, ,
\end{eqnarray}
to which we can relate the interdot Coulomb exchange interaction
\begin{eqnarray}
X_{0} = \langle L(1)L(2) | \hat{C} | R(1)R(2)\rangle = l^{2}U_{0} \,\, ,
\end{eqnarray}
and finally there is the matrix element
\begin{eqnarray}
w_{0} & = & \langle R(1)L(2) | \hat{C} | L/R(1) L/R(2) \rangle \nonumber \\
& = &   \frac{1}{\sqrt{2}} \langle S(1,1) |\hat{C}| S(2,0)/S(0,2) \rangle \nonumber \\
& = & l c\sqrt{b}e^{-\frac{d_0^2}{4b}} I_0 \left( \dfrac{d_0^2}{4b}
\right)
\end{eqnarray}

In the basis built of  the orthogonalized $\Phi_{L/R}$ states we have then the bare tunneling
\begin{eqnarray}
t' & = &  -\dfrac{d_0 \ l}{2 (1-l^2)} \Bigg(2d_0 -(d_0-f) \text{erf}[\sqrt{b} (d_0-f)] +2 f \text{\text{erf}}[\sqrt{b} f] \nonumber\\
& & -(d_0+f) \text{\text{erf}}[\sqrt{b}(d_0+f)] \nonumber\\
& & -\dfrac{e^{-b (d_0+f)^2} (1+e^{4 b d_0 f}-2 e^{b d_0 (d_0+2
f)})}{\sqrt{b \pi }} \Bigg) \,\, .
\end{eqnarray}

The matrix elements of the Coulomb interaction in this basis are given by
\begin{eqnarray}
w & = & \dfrac{2w_{0} - l ((1+l^2) U_{0}+D_{0}+E_{0}-2 lw_{0})} {2(1-l^2)^2} \,\, ,~~~~~~~\\
V_{S} & = & \frac{\left(l^2+l^4\right) U_{0}+D_{0}+  E_{0}-4lw_{0}}{\left(1-l^2\right)^2} \,\, , \\
V_{T} & = & \frac{D_{0}-E_{0}}{1-l^2} \,\, , \\
U & = & \frac{\left(2-l^2+l^4\right) U_{0}+ l^2 \left(D_{0}+E_{0} -4w_{0}/l \right)}{2 \left(1-l^2\right)^2} \,\, , \\
X & = &-\frac{l^2 \left[ \left(-3+l^2\right) U_{0}-D_{0}-E_{0}+4w_{0}/l \right]}{2 \left(1-l^2\right)^2} \,\, .
\end{eqnarray}

\end{document}